\documentstyle[12pt,aasms]{article}
\def\gapprox{\mathrel{\vcenter{\offinterlineskip \hbox{$>$}
    \kern 0.3ex \hbox{$\sim$}}}}
\def\lapprox{\mathrel{\vcenter{\offinterlineskip \hbox{$<$}
    \kern 0.3ex \hbox{$\sim$}}}}
\begin{document}
\setlength{\baselineskip}{12pt}

\title{Hydrodynamical Non-radiative Accretion Flows in Two-Dimensions}

\author{James M. Stone$^{1}$
James E. Pringle$^{1}$}
\affil{$^{1}$Institute of Astronomy, Cambridge University, Madingley Road,
Cambridge CB3 0HA, UK}
\affil{$^{2}$Department of Astronomy, University of Maryland, College Park,
MD 20742 USA}
\and
\author{Mitchell C. Begelman$^{1,2,3}$}
\affil{$^{3}$JILA, University of Colorado, Boulder, CO 80309-0440 USA}
\affil{$^{4}$ITP, University of California, Santa Barbara, CA 93106-4030 USA}

\begin{abstract}
Two-dimensional (axially symmetric) numerical hydrodynamical
calculations of accretion flows which cannot cool through emission of
radiation are presented.  The calculations begin from an equilibrium
configuration consisting of a thick torus with constant specific
angular momentum.  Accretion is induced by the addition of a small
anomalous azimuthal shear stress which is characterized by a
function $\nu$.  We study the flows generated as the amplitude and
form of $\nu$ are varied.  A spherical polar grid which spans more than
two orders of magnitude in radius is used to resolve the flow over a
wide range of spatial scales.  We find that convection in the inner
regions produces significant outward mass motions that carry away both
the energy liberated by, and a large fraction of the mass participating
in, the accretion flow.  Although the instantaneous structure of the
flow is complex and dominated by convective eddies, long time averages
of the dynamical variables show remarkable correspondence to
certain steady-state solutions.  The two-dimensional structure
of the time-averaged flow is marginally stable to the H{\o}iland
criterion, indicating that convection is efficient.  Near the
equatorial plane, the radial profiles of the time-averaged variables
are power-laws with an index that depends on the radial scaling of the
shear stress.  A stress in which $\nu \propto r^{1/2}$ recovers the
widely studied self-similar solution corresponding to an
``$\alpha$-disc".  We find that regardless of the adiabatic index of
the gas, or the form or magnitude of the shear stress, the mass inflow
rate is a strongly increasing function of radius, and is everywhere nearly
exactly balanced by mass outflow.  The net mass
accretion rate through the disc is only a fraction of the
rate at which mass is supplied to the inflow at large radii, and is
given by the local, viscous accretion rate associated with the flow
properties near the central object.

\end{abstract}


\section{Introduction}

There is considerable interest in accretion flows which cannot lose
internal energy through radiative cooling, since they may be relevant
to accretion of diffuse plasma onto compact objects such as black holes
and neutron stars.  Steady-state solutions for the
vertically-averaged radial structure of such flows have
been developed within the context of self-similarity (for example,
Ichimaru 1977; Begelman \& Meier 1982; Narayan \& Yi 1994; Abramowicz
et al. 1995) assuming angular momentum transport is mediated by an
anomalous shear ``viscosity".  Solutions in which the mass accretion
rate is constant with radius, and in which most of the gravitational
binding energy liberated by accretion is stored as thermal energy
(often called advection dominated accretion flows, or ADAFs)
have been the focus of many recent studies motivated, for example, by the
unusually low high-energy luminosity of some accreting black hole
candidates (Narayan et al. 1998).

There is, however, a long standing question as to whether
vertically-averaged solutions are indeed a good representation of the
true multidimensional flow.  For example, it is possible that the
energy liberated by accretion could drive an outflow in the polar
regions which accompanies accretion at the equator (Narayan \& Yi 1994;
1995;  Xu \& Chen 1997).  If the outflow carries a considerable
fraction of the mass, energy, and angular momentum available in the
accretion flow, it will have important consequences for the global
nature of the solution (for example, the mass inflow rate at the
equator can no longer be constant with radius).  Recently, steady-state
self-similar adiabatic inflow-outflow solutions (or ``ADIOS") have been
developed in generality by Blandford \& Begelman (1999a; 1999b,
hereafter BB).  Spectral models of such solutions have been calculated
by Quataert \& Narayan (1999).  Despite the greater generality of the
ADIOS models, there remains uncertainty as to how important properties of the
outflow such as the mass loss rate and terminal velocity are
determined.

Outflow from a non-radiative accretion flow will only be captured in a
multi-dimensional treatment of the problem; angle-averaged solutions
find only inflow in the inner regions (Ogilvie 1999).  In a study of
steady flows in thin discs, Urpin (1983; 1984) found that accretion at
the surfaces of the disc could be accompanied by outflow along the
equatorial (disc) plane.  Gilham (1981) considered two-dimensional
self-similar solutions which were separable in spherical polar
coordinates.  His solutions did not contain outflow, but they were also
convectively unstable and therefore could not be steady.

An alternative technique for studying multidimensional non-radiative
accretion flows is to use numerical methods to solve the time-dependent
hydrodynamical equations directly.  With such methods, a variety of
problems related to accretion onto black holes have been studied.  For
example, the formation of rotationally supported thick tori from
inviscid accretion of gas with various initial angular momentum
distributions has been reported (Hawley, Smarr \& Wilson 1984a; 1984b;
Hawley 1986; see also Molteni et al. 1994; Ryu et al. 1995; Chen et
al.  1997).  More recently, an accretion flow driven by an anomalous
``viscosity" around both non-rotating (Igumenshchev et al. 1996) and
rotating (Igumenshchev \& Beloborodov 1997) black holes has been
simulated, revealing a variety of interesting features.  The dynamic
range of spatial scales in these latter simulations is too small for a
direct comparison of the time-averaged state to the steady-state
self-similar solutions discussed above, though more recently these
authors have extended their calculations to include a much larger range
in radius (Igumenshchev 1999; Igumenshchev \& Abramowicz 1999,
hereafter collectively referred to as IA99).  These new calculations
reveal that bipolar outflows, strong convection, and quasi-periodic
variability can result as the strength of the viscosity is varied.

Contemporaneous to the latest work of Igumenshchev, we have performed
two-dimensional (axisymmetric) hydrodynamic calculations of
non-radiative accretion flows which span over two orders of magnitude
in radius.  Our calculations begin with a well defined initial
equilibrium state: a constant angular momentum ``thick" torus.  Because
ultimately the accretion is driven by an assumed and {\em ad hoc} shear
stress, we study several different forms and a wide range of amplitudes
for this stress.  We find that, for the models studied here, the
instantaneous flow is dominated by strong small-scale convection in the
inner regions.  Some of the time-averaged properties of the flow are remarkably
independent of the form of the shear stress, at least for the forms
adopted here.  In every case we find that at each radius, the mass flux
carried inward by convective eddies is nearly exactly balanced by outward
motions, so that the net mass {\em
accretion} rate is small compared to the local mass {\em inflow} rate.
For some forms of the anomalous stress, the regions of inflow and
outflow are distributed equally throughout the disc, while for others
the outflow occurs preferentially toward the poles, so that on the
largest scales the time-averaged flow forms a global circulation
pattern of inflow at the equator and slow outflow at the poles.  None
of our models show the production of powerful unbound winds.  Most
importantly, we find the net mass accretion rate through the disc is
determined locally by the properties of the flow near the surface of
the inner boundary; it is much smaller than the inflow
rate at large radii.

Although our calculations have been pursued independently of the work
of IA99, our formulation of the problem is very similar.  In fact we
differ only through the use of a different numerical method, a
different representation of the shear stress (see \S2.1), and different
initial conditions (see \S2.2).  It is likely that the only difference
of any importance is the representation of the shear stress.  We
compare our approach and results in more detail throughout this paper.

The paper is organized as follows.  In \S2 we describe our methods.
In \S3 we present the results of our hydrodynamical calculations.
In \S4 we discuss our results, while in \S5 we summarize and conclude.

\section{Method}

\subsection{The Equations of Motion}

To compute the models discussed here, we solve the equations of hydrodynamics
\begin{equation}
 \frac{d\rho}{dt} + \rho \nabla \cdot {\bf v} = 0,
\end{equation}
\begin{equation}
 \rho \frac{d{\bf v}}{dt} = -\nabla P - \rho \nabla \Phi + \nabla \cdot {\sf T},
\end{equation}
\begin{equation}
 \rho \frac{d(e/\rho)}{dt} = -P\nabla \cdot {\bf v} +{\sf T}^{2}/\mu,
\end{equation}
where $\rho$ is the mass density, $P$ the pressure, ${\bf v}$ the
velocity, $e$ the internal energy density, and ${\sf T}$ the anomalous
stress tensor.  The magnitude of the shear stress is determined by the
coefficient $\mu$; we discuss the forms for $\mu$ used here below.  The
$d/dt \equiv \partial/\partial t + {\bf v}\cdot\nabla$ denotes the
Lagrangian time derivative.  A strictly Newtonian gravitational
potential $\Phi$ for a central point mass $M$ is used here.  Using a
pseudo-Newtonian potential (Paczy\'{n}ski \& Wiita 1980) to approximate
general relativistic effects in the inner regions is in principle
straightforward but not of importance to the present investigation.  We
adopt an adiabatic equation of state $P=(\gamma -1)e$, and consider
models with both $\gamma =5/3$ and $4/3$ (the latter value may provide
a better, albeit crude, representation of the dynamics of a
radiation dominated fluid).  These equations are solved in spherical
polar coordinates $(r,\theta,\phi)$.

The final term on the RHS of equations 2 and 3 represent anomalous
shear stress and heating respectively added to drive angular momentum
transport and accretion.  It must be emphasized that the shear stress we
wish to approximate is not the result of a true kinematic viscosity.
In reality, we expect Maxwell stresses associated with MHD turbulence
driven by the magnetorotational instability (MRI) to
provide angular momentum transport in accretion flows (Balbus \& Hawley 1998).  Modeling such
processes from first principles requires fully three-dimensional MHD
calculations:  an important and necessary step for future
investigations but beyond the scope of this paper.  To approximate the
effects of magnetic stresses, we assume the azimuthal components of the
shear tensor ${\sf T}$ are non-zero and, in spherical polar coordinates, are
given by
\begin{equation}
  T_{r\phi} = \mu r \frac{\partial}{\partial r}
	\left( \frac{v_{\phi}}{r} \right),
\end{equation}
\begin{equation}
  T_{\theta\phi} = \frac{\mu \sin \theta}{r} \frac{\partial}{\partial
  \theta} \left( \frac{v_{\phi}}{\sin \theta} \right) .
\end{equation}
This form is similar to the shear stress in a viscous fluid, in which
case $\mu = \nu \rho$ is the coefficient of shear viscosity, and $\nu$
the kinematic viscosity coefficient.   We emphasize, however, that
unlike a viscous fluid, we have assumed the non-azimuthal components of
the shear stress are zero, because the MRI is driven only by the shear
associated with the orbital dynamics, and therefore we do not expect it
to affect poloidal shear to the same degree.  Local three-dimensional
MHD simulations of the properties of the MRI in the non-linear regime
confirm that the azimuthal components of the Maxwell stress are indeed
more than an order of magnitude larger than the poloidal components
(e.g. Hawley, Gammie, \& Balbus 1995; 1996; Brandenburg et al. 1995;
Stone et al. 1996).  IA99 use a shear stress which has a similar form
to equations 4 and 5, except they also include the poloidal stress
terms.

Although the local properties of the MRI in the nonlinear regime have
been well studied, the radial scaling of the Maxwell stresses in a
global MHD accretion flow is uncertain.  Thus, we adopt several
empirical forms for $\nu$ to study if and how the flow is changed.  In
particular, we consider models with (1) $\nu \propto \rho$, (2) $\nu =$
constant, and (3) $\nu \propto r^{1/2}$.  The first form is adopted
primarily as a numerical convenience; it ensures the stresses are
confined mostly to the torus where the density is large.  In fact, as
shown below, it results in a flow with constant density in the
equatorial plane, which implies the shear stress per unit volume is
nearly constant, i.e. it implies the the solutions for the first and
second forms of the stress should be similar.  The third form for the
shear stress corresponds to the expectations of an ``$\alpha$-disc"
model.  In the last two cases, the stress per volume is independent of
all dynamical variables, thus the flow cannot adjust itself to change
${\sf T}$.  Thus, we might expect models 2 and 3 to give quite
different accretion solutions.

It is important to question whether approximating angular momentum
transport in this manner rather than undertaking the requisite MHD
calculations is worthwhile.  Because ``viscous" non-radiative accretion
flows have been so widely discussed in the literature, it is useful to
calculate their time-dependent hydrodynamical evolution over a large
spatial scale both for comparison to studies of steady-state flows, and
to serve as a baseline for future time-dependent MHD calculations.
Because we find that the time-averaged properties of the flow are
remarkably independent of the form of the shear stress, it is possible
that some of the results found here will not be qualitatively different
in an MHD calculation, provided the angular momentum transport is
dominated by internal Maxwell stress (rather than external torques
provided by global fields).  Clearly MHD
calculations are warranted to investigate these issues.

\subsection{Initial Conditions}

Our simulations begin with an equilibrium state consisting of a
constant angular momentum torus embedded in a non-rotating,
low-density ambient medium in hydrostatic equilibrium.  The pressure
and density in the torus initially are related through a polytropic
equation of state $P=A\rho^{\gamma}$, so that the equilibrium structure
of the torus is given by (Papaloizou \& Pringle 1984)
\begin{equation}
 \frac{P}{\rho} = \frac{GM}{(n+1)R_{0}} \left[ \frac{R_{0}}{r} - \frac{1}{2}
\left( \frac{R_{0}}{r \sin \theta} \right)^{2} - \frac{1}{2d} \right] .
\end{equation}
Here $n=(\gamma-1)^{-1}$ is the polytropic index, $R_{0}$ is the radius
of the center (density maximum) of the torus, and $d$ is a distortion
parameter which measures the shape of the cross-sectional area  
of the torus.  Physical solutions to equation 6 require $d>1$, with
values near one giving small, nearly circular cross sections.  Without loss
of generality, we choose units such that $G=M=R_{0}=1$, so that the
maximum density is
\begin{equation}
 \rho_{max} = \left(\frac{1}{(n+1)A} \left[\frac{d-1}{2d} \right] \right)^{n}.
\end{equation}
For given values of $\rho_{max}$, $d$ and $n$, equation 7 determines
$A$.  In most cases we adopt $\rho_{max}=1$, although in some models we
specify the mass contained in the torus instead.

The ambient medium in which the torus is embedded has density
$\rho_{0}$ and pressure $P_{0} = \rho_{0}/r$.  The mass and pressure of
the ambient medium should be negligibly small; here we choose
$\rho_{0}=10^{-4}$ so that the ratio of the mass in the ambient medium
in $r<R_{0}$ to that in the torus $M_{t}$ is $\sim 4\times10^{-4}$.
Similarly, the ratio of the maximum pressure in the torus to the
ambient pressure at $r=R_{0}$ is $\sim 5\times10^{-3}$.  In order to
guarantee an exact numerical equilibrium initially, we first initialize
the density everywhere, and then integrate the radial equation of motion
inwards from the outer boundary to compute the pressure using the
numerical difference formula in our hydrodynamics code.

\subsection{Numerical Methods}

All of the calculations presented here use the ZEUS-2D code described
by Stone \& Norman (1992).  No modifications to the code were required,
apart from the addition of the shear stress terms.  These terms are
updated in an operator split fashion separately from the rest of the dynamical
equations.   For stability, these terms must be integrated using a time
step which satisfies $\bigtriangleup t < \min (\bigtriangleup r,
r\bigtriangleup \theta)^2 / \nu$, where $\bigtriangleup r$ and
$\bigtriangleup \theta$ are the radial and angular grid spacing, and the
minimum is taken over the entire mesh.  Whenever this requirement is
smaller than the time step used for the hydrodynamic equations, we
sub-cycle: that is we integrate the viscous terms repeatedly at the
smaller timestep until one hydrodynamical timestep has elapsed.

Our computational grid extends from an inner boundary at $r=R_{B}$ to
$4R_{0}$.  In most simulations, we choose $R_{B}=0.01R_{0}$, giving a
spatial dynamic range over which an accretion flow can be established
of two orders of magnitude.  In some models we set $R_{B}=0.005R_{0}$
with the outer boundary at $800R_{B}$, so that the grid extends nearly
three orders of magnitude in radius. Such large dynamic ranges are
essential for investigating an accretion flow in the self-similar
regime, but are challenging because of the wide range in dynamical
(orbital) timescales.  For example, for the standard numerical
resolution given below, one orbit of evolution at $r=R_{0}$ requires
10-15 hours on currently available workstations, while calculations at
twice the standard resolution take more than 10 times longer.

In order to adequately resolve the flow over such a large spatial
scale, it is necessary to adopt a non-uniform grid.  We choose a
logarithmic grid in which $(\bigtriangleup r)_{i+1} / (\bigtriangleup
r)_{i} = \sqrt[N_{r}-1]{10}$.  This gives a grid in which
$\bigtriangleup r \propto r$, with $N_{r}$ grid points per decade in
radius.  Similarly, in order to better resolve the flow at the
equator, we adopt non-uniform angular zones with $(\bigtriangleup
\theta)_{j} / (\bigtriangleup \theta)_{j+1} = \sqrt[N_{\theta}-1]{4}$
for $0 \leq \theta \leq \pi/2$ (that is the zone spacing is decreasing
in this region), and $(\bigtriangleup \theta)_{j+1} / (\bigtriangleup
\theta)_{j} = \sqrt[N_{\theta}-1]{4}$ for $\pi/2 \leq \theta \leq
\pi$.  This gives a refinement by a factor of four in the angular grid
zones between the poles and equator.  Equatorial symmetry is not
assumed.  Our standard resolution is $N_{r}=64$ and $N_{\theta}=44$,
giving a grid with a total size  of $168\times88$ zones.  We have also
computed a high resolution model with $N_{r}=128$ and $N_{\theta}=80$,
giving a grid with a total size  of $334\times160$ zones.

We adopt outflow boundary conditions (projection of all dynamical
variables) at both the inner and outer
radial boundaries.  We further set $T_{r\phi}=0$ at the inner radial 
boundary.  In the angular direction, the boundary conditions are set
by symmetry at the poles (including setting $T_{\theta\phi}=0$).

In some numerical simulations, numerical transport of angular momentum
can limit the accuracy of the results.  In fact, the consistent
transport algorithms implemented in the ZEUS-2D code (Stone \& Norman
1992) are specifically designed to reduce this effect.  As a test, we
have evolved a torus with $\nu=0$ for 2 orbits at $r=R_{0}$.  Apart
from a slight smoothing of the sharp edges of the torus over 2-3 grid
zones, we find no evolution.  The angular momentum of the torus is
conserved to one part in $10^{6}$, and the total rotational kinetic
energy to one part in $10^{3}$.

We have also tested our implementation of the shear stress terms by
comparing the evolution of the angle-integrated surface density at
early times with the exact analytic solution for a thin viscous torus (e.g.
Figure 1 of Pringle 1981).  We find excellent correspondence between
our numerical and the analytic solution up to a time of roughly 0.8
orbits, beyond which effects associated with the multidimensional nature
of the flow in our simulations begin to become important.

\section{Results}

\subsection{A Fiducial Model}

Table 1 summarizes the properties of the simulations discussed here.
Columns two through six give the coefficient of the shear stress $\nu$,
numerical resolution, radial extent of the numerical grid expressed as
the ratio $R_{0}/R_{B}$, distortion parameter $d$ of the torus, and
adiabatic index $\gamma$ respectively.  The final two columns are the
final time $t_{f}$ at which each simulation is stopped (all times in
this paper are reported in units of the orbital time at $r=R_{0}$), and
the time-averaged mass accretion rate through the inner boundary
measured near the end of the simulation,
in units of the initial mass of the torus $M_{t}$ per orbit.

\begin{table*}
\footnotesize
\begin{center}
\caption{Properties of Simulations}
\begin{tabular}{cccccccc} \\ \hline 
Run 
& $\nu$
& Resolution$^{(a)}$
& $R_{0}/R_{B}$
& $d$
& $\gamma$
& $t_{f}$ (orbits$^{(b)}$)
& $\dot{M}_{\rm acc}(R_{B})^{(c)}$ $(M_{t}/$orbit$^{(b)}$) \\ \hline

A & $10^{-2}\rho$    &  64 & 100 & 1.125 & 5/3 & 4.325 & $1.0\times10^{-3}$ \\
B & $10^{-2}\rho$    & 128 & 100 & 1.125 & 5/3 & 2.274 & $8.2\times10^{-4}$ \\
C & $10^{-2}\rho$    &  64 & 200 & 1.125 & 5/3 & 1.947 & $5.0\times10^{-4}$ \\
D & $10^{-2}\rho$    &  64 &  50 & 1.125 & 5/3 & 4.000 & $1.8\times10^{-3}$ \\
E & $10^{-2}\rho$    &  64 & 100 & 1.125 & 4/3 & 2.155 & $1.6\times10^{-3}$ \\
F & $5\times10^{-3}\rho$&64& 100 & 1.042 & 5/3 & 3.539 & $9.1\times10^{-4}$ \\
G & $5\times10^{-3}\rho$&64& 100 & 1.042 & 4/3 & 1.391 & $1.9\times10^{-3}$ \\
H & $10^{-3}\rho$    &  64 & 100 & 1.125 & 5/3 &20.389 & $6.6\times10^{-5}$ \\
I & $10^{-1}\rho$    &  64 & 100 & 1.125 & 5/3 & 0.593 & $3.0\times10^{-3}$ \\
I$^{\prime(d)}$& $10^{-1}\rho$&64&100&1.125&5/3& 3.110 & $2.4\times10^{-3}$ \\
J & $10^{-3}$        &  64 & 100 & 1.125 & 5/3 & 5.479 & $8.3\times10^{-4}$ \\
K & $10^{-3}r^{1/2}$ &  64 & 100 & 1.125 & 5/3 & 13.6181 & $1.1\times10^{-3}$ \\
\hline
\end{tabular}\\
\end{center}
(a) grid points per decade in radius, (b) at $r=R_{0}$,
(c) averaged over last orbit, (d) restarted from Run A\\
\end{table*}

We find models in which $\nu \propto \rho$ are the quickest to settle
into a steady state; we discuss these models in detail to start.  Runs
A and B are low- and high-resolution simulations of a fiducial model
with $\nu = 10^{-2} \rho$.  Figure~1 plots the time evolution of the
density in Run B.  In each panel, gray shading is used to denote
regions of the torus in which $v_{r} > 0$.  Initially, the evolution is
as predicted (Pringle 1981): most of the mass loses angular momentum
and falls inward forming a wedge-shaped structure, while a small
fraction of the mass gains angular momentum and moves outward.  In
fact, as indicated by the shading, the surface layers of the inner
regions actually move outward -- they form a Kelvin-Helmholtz (K-H)
roll at the surface of the torus by $t=0.5$.  By $t=1$ material has
begun to accrete through the inner boundary, while outflow along the
surface of the torus has carried the first K-H roll out to $r=1.75$,
and a second roll-up is forming at $r=1$.  Thereafter, the central
wedge-shaped flow thickens considerably, so that by the end of the
calculation the density contours in the inner regions are nearly
radial.  The wiggles in the contours in this region are caused by large
amplitude density fluctuations in the flow associated with strong
convective motions occurring at small radii.  This convection drives a
low-velocity outflow from the inner regions, in fact, all the material
located at angles $\theta \gapprox 45$ degrees from the equator is
outflowing.  This flow remains bound, however, so that it eventually
cascades back onto the surface of the torus and merges with it in the
outer regions.  The radial pattern of shading at $t=2.274$ in the inner
regions, indicative of alternating columns of material with opposite
sign in $v_{r}$, is another sign of large-scale convective flows.

\begin{figure*}
\begin{picture}(504,360)
\put(0,0){\includegraphics{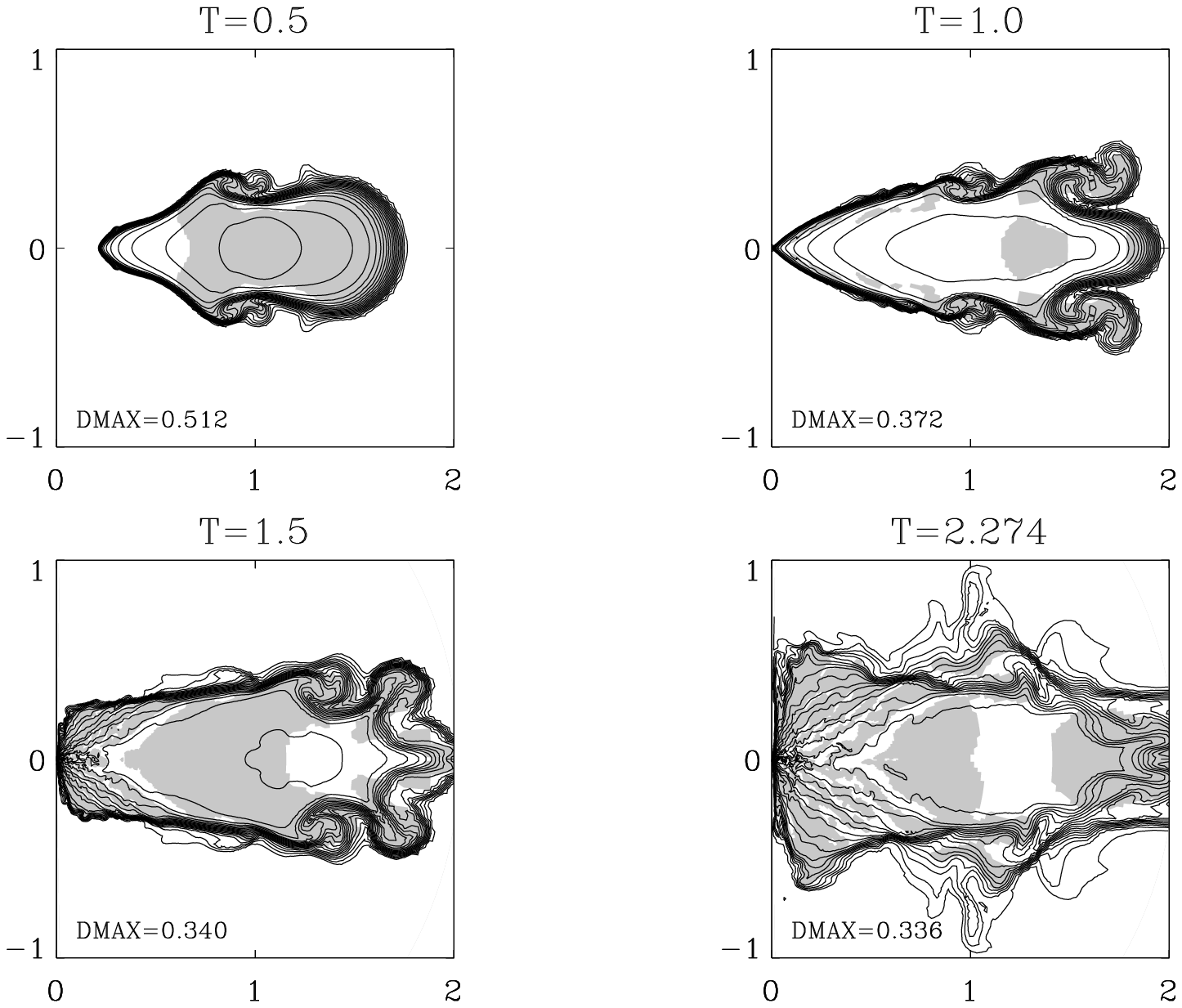}}
\end{picture}
\caption{Evolution of the density in a fiducial model (Run B).  Time is measured
in units of the orbital time at $r=1$.  Twenty logarithmic contours are
used between the density maximum (given in each panel) and $10^{-4}$.
The shaded regions have $v_{r}>0$.}
\end{figure*}

Figure~2 is a plot of the time-history of the angle-integrated mass
accretion rate $\dot{M}_{\rm acc}$ at $r=0.01$ (i.e. the inner boundary),
0.05, and 0.1 in Run A, and at $r=0.01$ in Run B, where
\begin{equation}
  \dot{M}_{\rm acc}(r) =2\pi r^{2}\int_{0}^{\pi} \rho v_{r} \sin\theta d\theta.
\end{equation}
Because of the computational expense,
Run B has not been evolved beyond 2.274 orbits.  Note that the mass
accretion rate through the inner boundary converges on the same value
at both resolutions.  Moreover, the time-averaged accretion rate is
independent of radius at late times, indicating the flow has reached a
steady state over at least one decade in radius.  Because the plot
shows the {\em integrated} accretion rate, it does not reveal the
magnitude of the inflow near the equator or the outflow near the poles,
but only the difference.  This net mass accretion rate is very small,
it would take $M_{t}/\dot{M} \approx 10^{3}$ orbits to accrete the
entire torus.  (Of course some material would never be accreted as required
by energy and angular momentum conservation.)

\begin{figure*}
\begin{picture}(252,180)
\put(0,0){\includegraphics{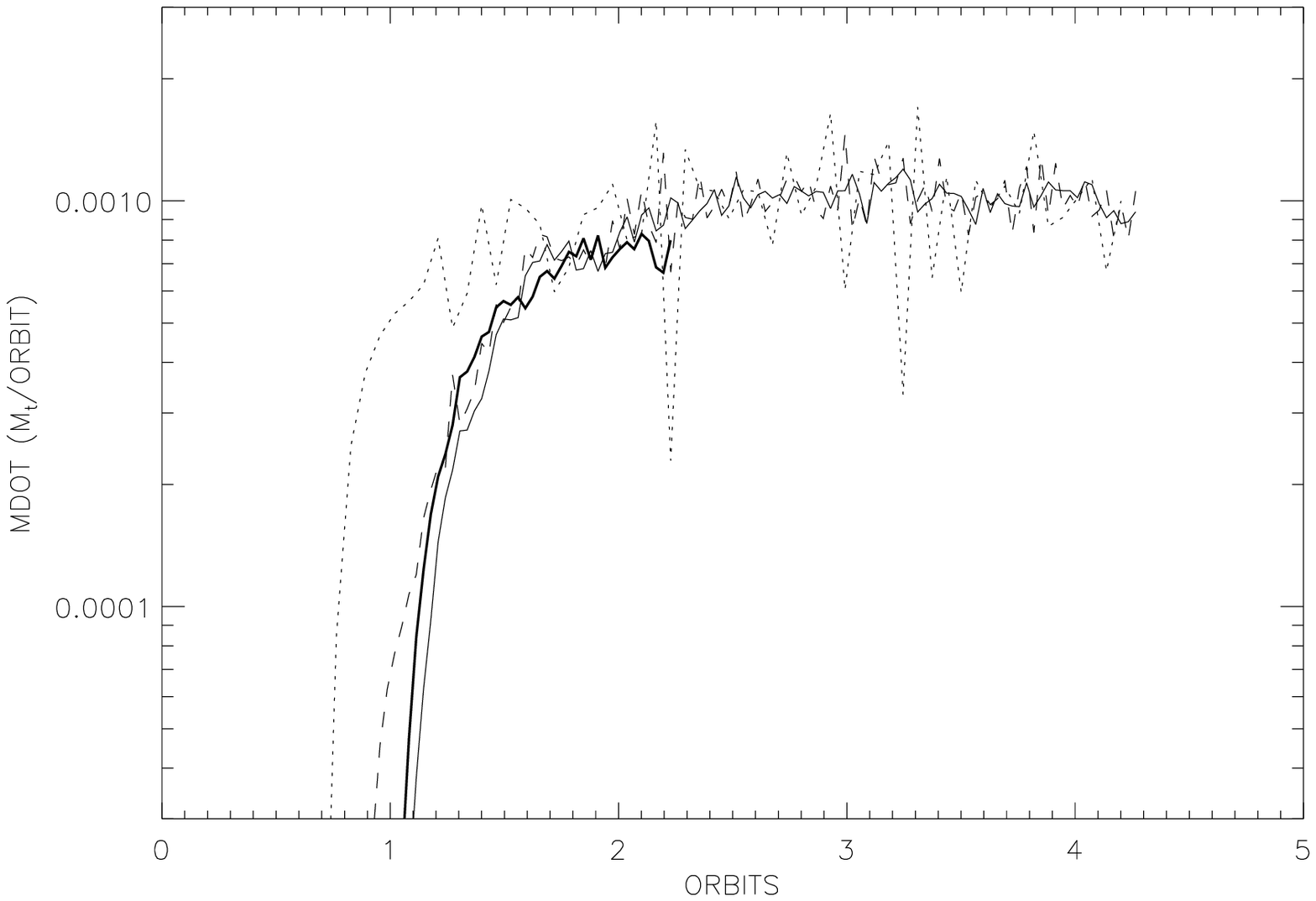}}
\end{picture}
\caption{Total mass accretion rate at $r=R_{B}$ (solid line),
$r=5R_{B}$ (dashed line), and $r=10R_{B}$ (dotted line) in the
fiducial model (Run A).  The bold solid line is the total mass
accretion rate at $r=R_{B}$ in the high resolution simulation (Run B).}
\end{figure*}

To reveal the complex structure of the flow at small radii, in Figure~3
we plot snapshots of the density, entropy (for a non-radiating fluid)
$S=\ln (P/\rho^{\gamma})$,
and angular momentum excess compared to Keplerian motion $\delta l =
\rho(v_{\phi}r\sin \theta - \sqrt{r\sin \theta})$ at $t=1.95$ for the
central region $r<0.1$.  The complexity of the flow on these scales is
evident.  Overall, the density is strongly stratified from the poles to
the equator, however on small scales large amplitude fluctuations
associated with convection dominate the image.  Several regions of
dense blobs and filaments (which move inward), and low density bubbles
(which move outward) are clearly evident.  Note that at some radii, the
angular position of the density maximum is displaced far from the
equator, an indication of how strongly the convective motions dominate
the flow.  The entropy plot shows the generic result for a
non-radiative accretion flow that $S$ is maximum along the poles (where
the density is a minimum), and smallest along the equator.  However,
once again the dominant pattern in the image are the bubbles and
filaments associated with convection.  Finally, the plot of $\delta l$
shows that overall, the flow near the poles has an excess of angular
momentum compared to Keplerian, while near the equator large amplitude
fluctuations of both excess and deficit are present.  A plot of the
instantaneous radial velocity on these scales is complex, revealing
that both inflow and outflow occurs at some place along most radial slices.
However, there is an overall pattern to the flow: inflow dominates
near the equator and outflow near the poles.  Comparison of
the three images demonstrates that, as one would expect, regions of
highest density (which are observed to move inward) have low $S$ and a
deficit of angular momentum, while bubbles of low density (which are
observed to move outwards) have high $S$ and an excess of angular
momentum.  A detailed comparison of the properties of convection
in a rapidly rotating flow such as being studied here with the known
properties of stellar convection (e.g. Mestel 1999) would be fruitful,
but will not be discussed further here.

\begin{figure*}
\begin{picture}(510,342)
\put(0,0){\includegraphics{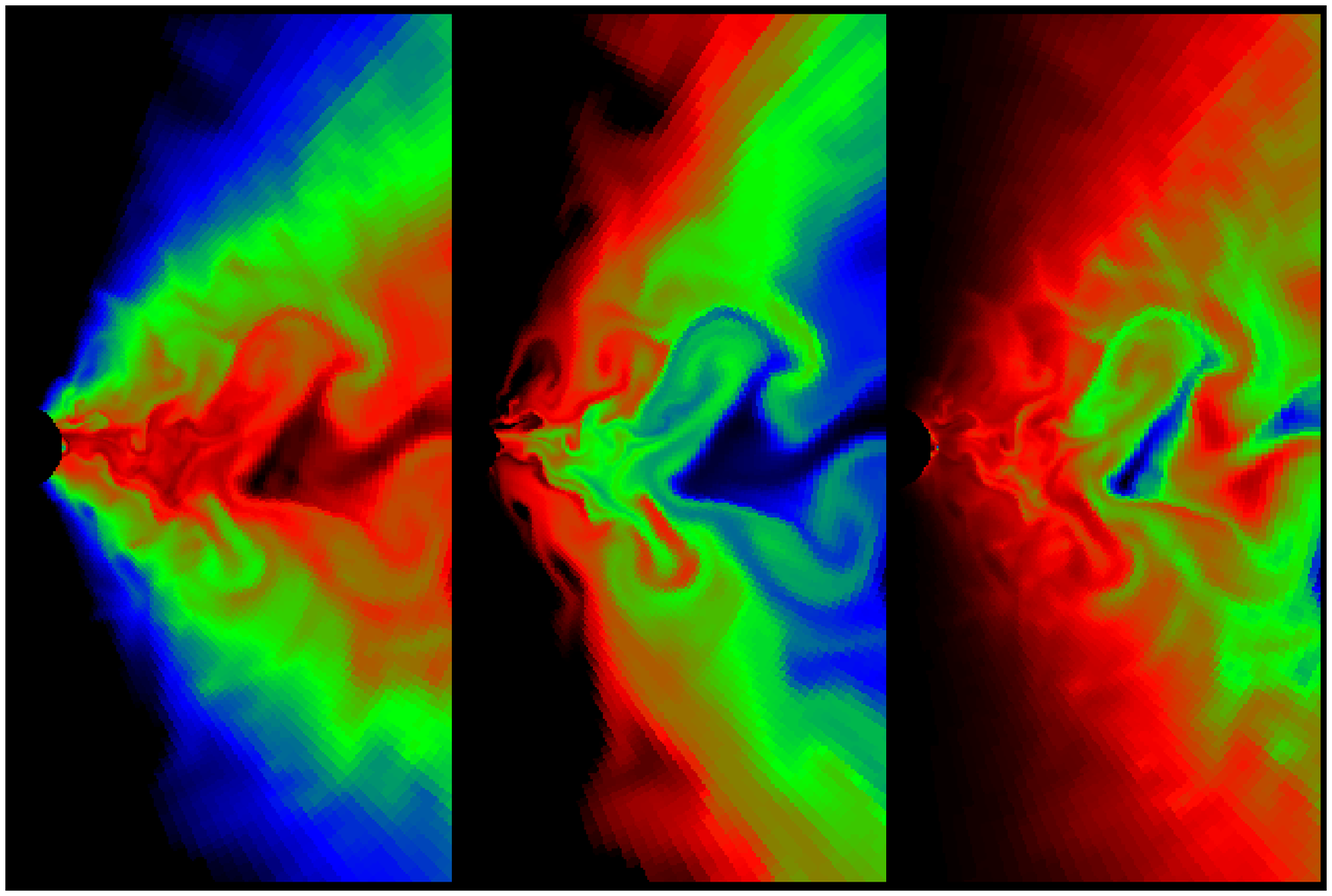}}
\end{picture}
\caption{Images of the logarithm of the density (left panel), specific
entropy (middle panel), and angular momentum excess compared to
Keplerian rotation (right panel) at $t=1.95$ orbits in the inner region
$r<0.1$ of Run B.  The colour table runs from black to blue through red to
black.  Near the equatorial plane, note the correlation between regions
of high (low) density, low (high) entropy, and a deficit (excess) of
angular momentum.}
\end{figure*}

In order to investigate the time-averaged properties of the flow, in
Figure~4 we plot contours of a number of time-averaged quantities,
including the specific angular momentum $L=v_{\phi}r\sin \theta$, and
the Bernoulli function $B \equiv v^{2}/2 + \gamma P/(\gamma-1)\rho -r^{-1}$.
The latter is shaded in regions where $B < 0$.  These averages are
constructed from 139 equally spaced data files between orbits 2.0 and
2.278.  The time-averaged variables form remarkably simple patterns
compared to the instantaneous snapshots (c.f. Fig. 3).  In each case
the contours form smooth curved surfaces which are symmetric with
respect to the equator.   The ordering of the variables according to
the shape of their contours (from concave to convex with respect to an
outward radial norm) is $\rho$, $P$, r, $\Omega$, $R$ $L$, $B$, and
$S$, where $r$ and $R$ are spherical and cylindrical radial coordinates
respectively.  Contours of the last three variables are in fact virtually
parallel, except for contours of $B<0$.  (Note in the region shown the
contours of $\rho$ are nearly radial, however since they must close at large
$r$ we consider this as an extreme limit of concavity.) 
This ordering is precisely that expected for stability
against the H{\o}iland criterion (Begelman \& Meier 1982; BB), thus we
conclude the mean state is marginally stable.  The convective motions
are therefore so efficient as to drive the mean flow to marginal
stability (since the instantaneous structure is clearly {\em
unstable}).  We note the shading of $B$ indicates that in a
time-average sense, $B<0$ for material near the equator (where there is
inflow), while $B>0$ nearer the poles (where there is outflow).  The
instantaneous structure of $B$ may contain large bubbles of positive
value embedded in negative values near the midplane.

\begin{figure*}
\begin{picture}(504,360)
\put(0,0){\includegraphics{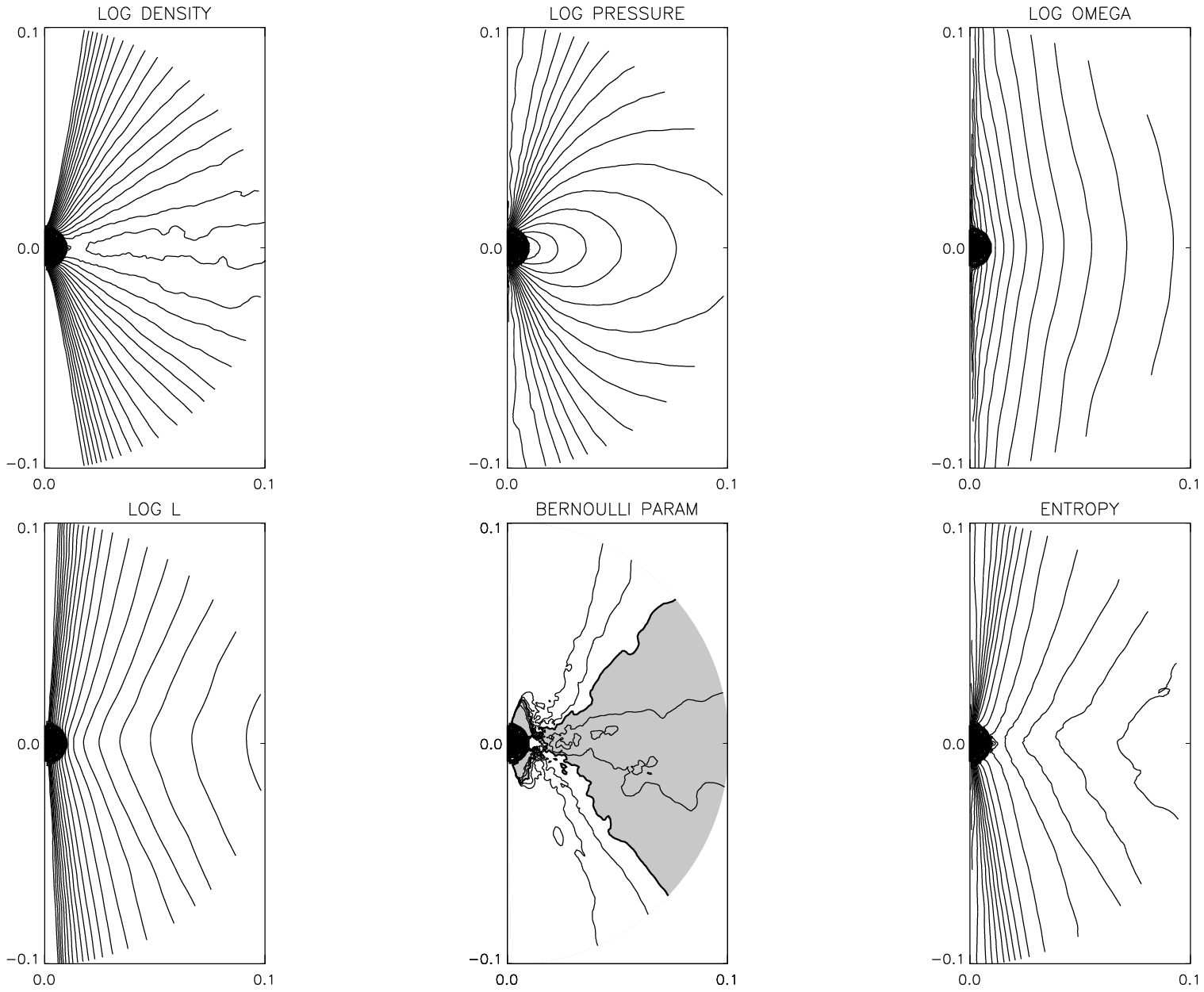}}
\end{picture}
\caption{Two-dimensional structure of a variety of time-averaged
quantities from the fiducial model (Run B).  The minimum and maximum of
$\log \rho$ are -3 and -0.69, of $\log P$ are -2.5 and 0.63,
of $\log \Omega$ are 1.4 and 4.2, of $\log L$ are -2 and -0.55
and of $S$ are 0.71 and 4.9.  For these variables, twenty equally spaced
contour levels are used.  Only five contours between -2 and 2 are
used for the plot of $B$ (although the minimum and maximum are -29 and 
$1.7\times10^{2}$ respectively), with shaded regions denoting $B<0$.}
\end{figure*}

Figure~5 plots the radial structure of the time-averaged flow near the
equatorial plane in Run B, averaged from $\theta = 84$ to 96 degrees.  The
time-averaging used to construct the plot is identical to that used in
Fig. 4.  It is clear that each variable in the plot can be described by
a radial power law, with $\rho \propto r^{0}$, $P \propto r^{-1}$,
$v_{\phi} \propto r^{-1/2}$, and $v_{r} \propto r^{-1}$.
Note the rotational velocity never falls below
$0.9$ times the Keplerian value, except directly next to the inner boundary.
The power-law index of $P$ is somewhat
uncertain: $P \propto r^{-1}$ is really only a good fit for $r<0.1$.
If we adopt the ansatz that the Bernoulli function is zero at all radii near
the equatorial plane (justified by the results shown in Fig. 4),
and that the kinetic energy is dominated by the
rotational velocity (which is nearly Keplerian), one would predict
\begin{equation}
  \gamma P / \rho = \frac{1}{3r}
\end{equation}
for $\gamma=5/3$.  In fact, equation 9 provides an approximate fit to
the radial profile of the time-averaged $C_{s}^{2} \equiv \gamma P / \rho$.

The result that $v_{r} \propto r^{-1}$ indicates the solution is not
strictly self-similar.  Since at $r=1$ the ratio $v_{r}/v_{\phi} \sim
10^{-2.5}$, the solution implies (if continued inward) that $v_{r} \sim
v_{\phi}$ at $r \sim 10^{-5}$.  We expect at this point the flow will
change qualitatively.  The particular scaling of the radial velocity
with radius in this flow undoubtedly is related to the form of the
shear stress used in this model, $\nu \propto \rho \propto r^{0}$.  We
note that since (from Fig. 5) the ratio $C_{s}^{2}/\Omega \approx 0.3r^{1/2}$,
it is not possible to characterize the shear stress by a constant
$\alpha_{eff} = \nu \Omega/C_{s}^{2}$.  Instead, $\alpha_{eff} \approx
10^{-2}r^{-1/2}$.

\begin{figure*}
\begin{picture}(504,360)
\put(0,0){\includegraphics{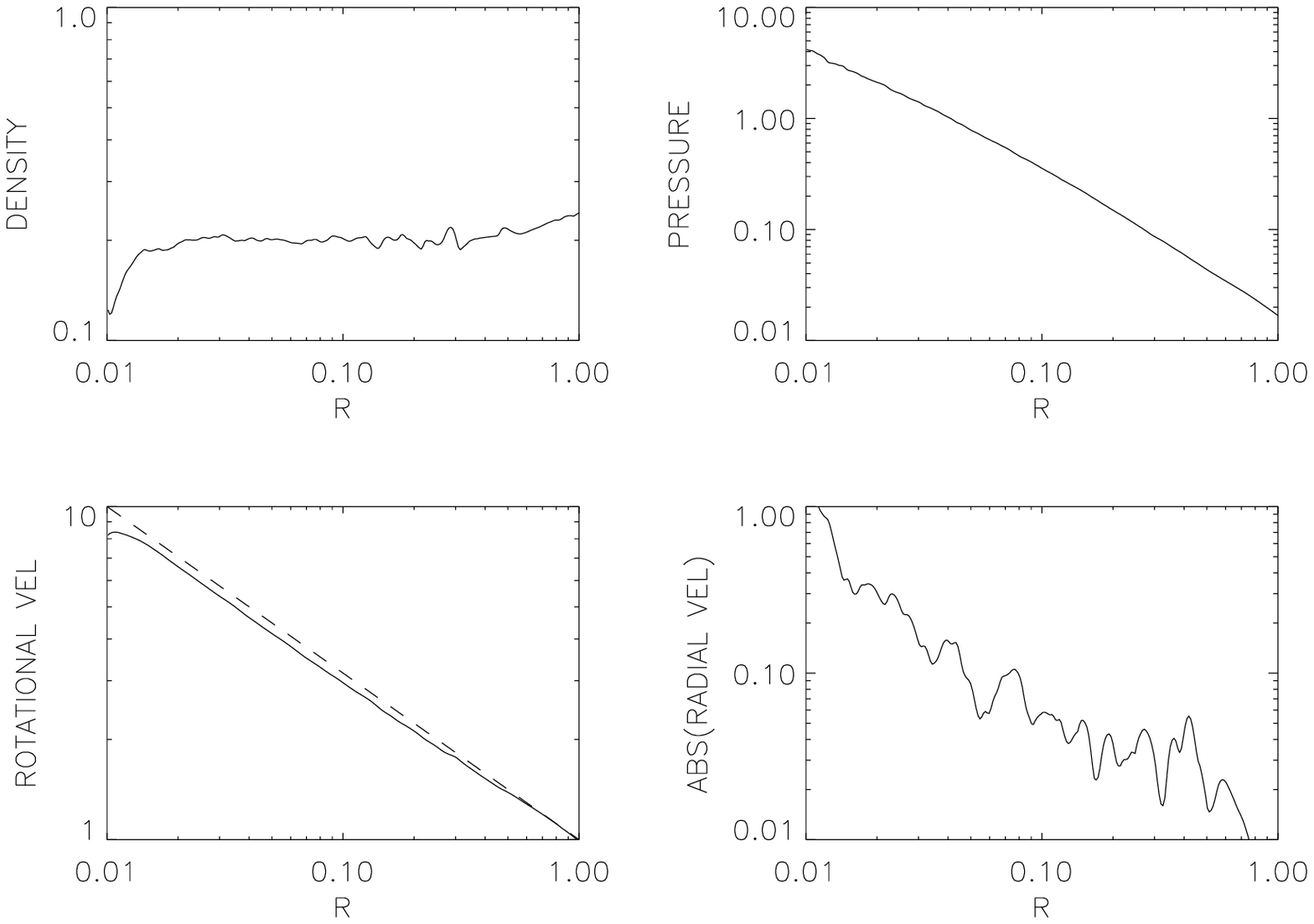}}
\end{picture}
\caption{Radial scaling of some time-averaged
quantities from the fiducial model (Run B).  The solution is averaged over
angle between $\theta=84$ and 96 degrees to construct each plot.
The dashed line in the plot of $v_{\phi}$ denotes Keplerian rotation
at the equator.}
\end{figure*}

\begin{figure*}
\begin{picture}(504,360)
\put(0,0){\includegraphics{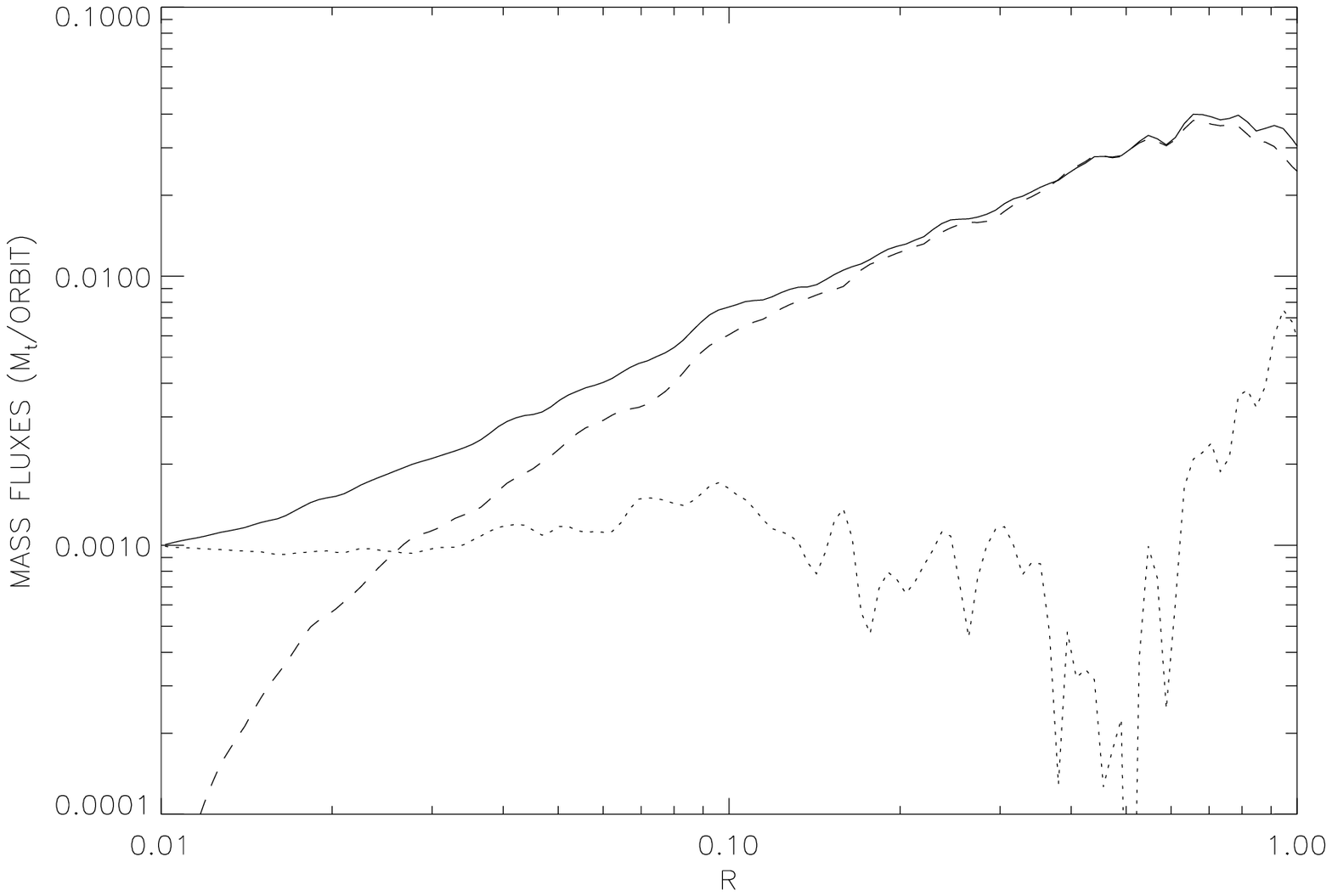}}
\end{picture}
\caption{Radial scaling of the total mass inflow rate $\dot{M}_{\rm in}$
(solid line), total mass outflow rate $\dot{M}_{\rm out}$ (dashed line), and
the net mass accretion rate  $\dot{M}_{\rm acc}$ (dotted line)
in the fiducial model (Run B).  These fluxes are defined by 
equations 8, 10, and 11 in the text.}
\end{figure*}

Figure 6 plots the radial structure of the time-averaged and
angle-integrated mass inflow and outflow rates ($\dot{M}_{\rm in}$
and $\dot{M}_{\rm out}$ respectively), defined as
\begin{equation}
 \dot{M}_{\rm in}(r) = 2\pi r^{2} \int_{0}^{\pi} \rho \min(v_{r},0)
   \sin \theta d\theta
\end{equation}
\begin{equation}
 \dot{M}_{\rm out}(r) = 2\pi r^{2} \int_{0}^{\pi} \rho \max(v_{r},0)
    \sin \theta d\theta 
\end{equation}
as well as the net mass accretion rate $\dot{M}_{\rm acc}$ defined in equation 8.
Clearly $\dot{M}_{\rm acc}=\dot{M}_{\rm out}+\dot{M}_{\rm in}$.  To
construct the plot, we time average the integral
(rather than integrating the time-averages) using the
same data as in Fig. 4.  It is evident that both $\dot{M}_{\rm in}$ and
$\dot{M}_{\rm out}$ increase roughly in proportion to $r$.  However, the
difference between the two ($\dot{M}_{\rm acc}$) is constant
with radius and is roughly equal to the local mass accretion rate
expected for a thin, viscous disc calculated using the properties of
the flow near the inner boundary, that is $\dot{M}_{\rm acc} \approx \pi
\nu(R_B) \Sigma(R_B)$.  Using the known form of the stress $\nu = 10^{-2}\rho$,
and the result from our simulations (see Figs. 5 and 7) that $\rho
\approx 0.2\sin \theta$ to compute $\Sigma(R_B)$ gives a viscous mass
accretion rate which is
within a factor of two of the measured $\dot{M}_{\rm acc}$.

Figure~7 plots the angular structure of the time-averaged flow at
radial positions of $r=0.02$, 0.05, 0.1, and 0.2, again using the same
time-averaging as in Fig. 4.  The angular structure of both the density
and pressure shows smooth variation from the poles to equator, with the
amplitude of $\rho$ virtually identical at each radius and the
amplitude of $P$ dropping as $r^{-1}$.  The radial velocity is
negative over most $\theta$ at $r=2R_{B}$.  At larger $r$ there are
still large variations in $v_{r}$, despite the large amount of data
used to construct the time-average.  The average $v_{r}$ is slightly
negative within about 30 degrees of the equator, and strongly positive
within 50 degrees of the poles.  The rotational velocity $v_{\phi}$ is
nearly constant with $\theta$ at every radial position, so that $L$
varies as $\sin \theta$.  From the plots it is difficult to distinguish
a surface that might be used to define the accretion disc as opposed to
the polar outflow.  Crudely, a surface might be defined near
$\theta=45$ degrees where the time-averaged radial velocity changes
sign.  Alternatively, the surface of $B=0$ in the fifth panel of Fig. 4
might be considered to define a disc.

\begin{figure*}
\begin{picture}(504,360)
\put(0,0){\includegraphics{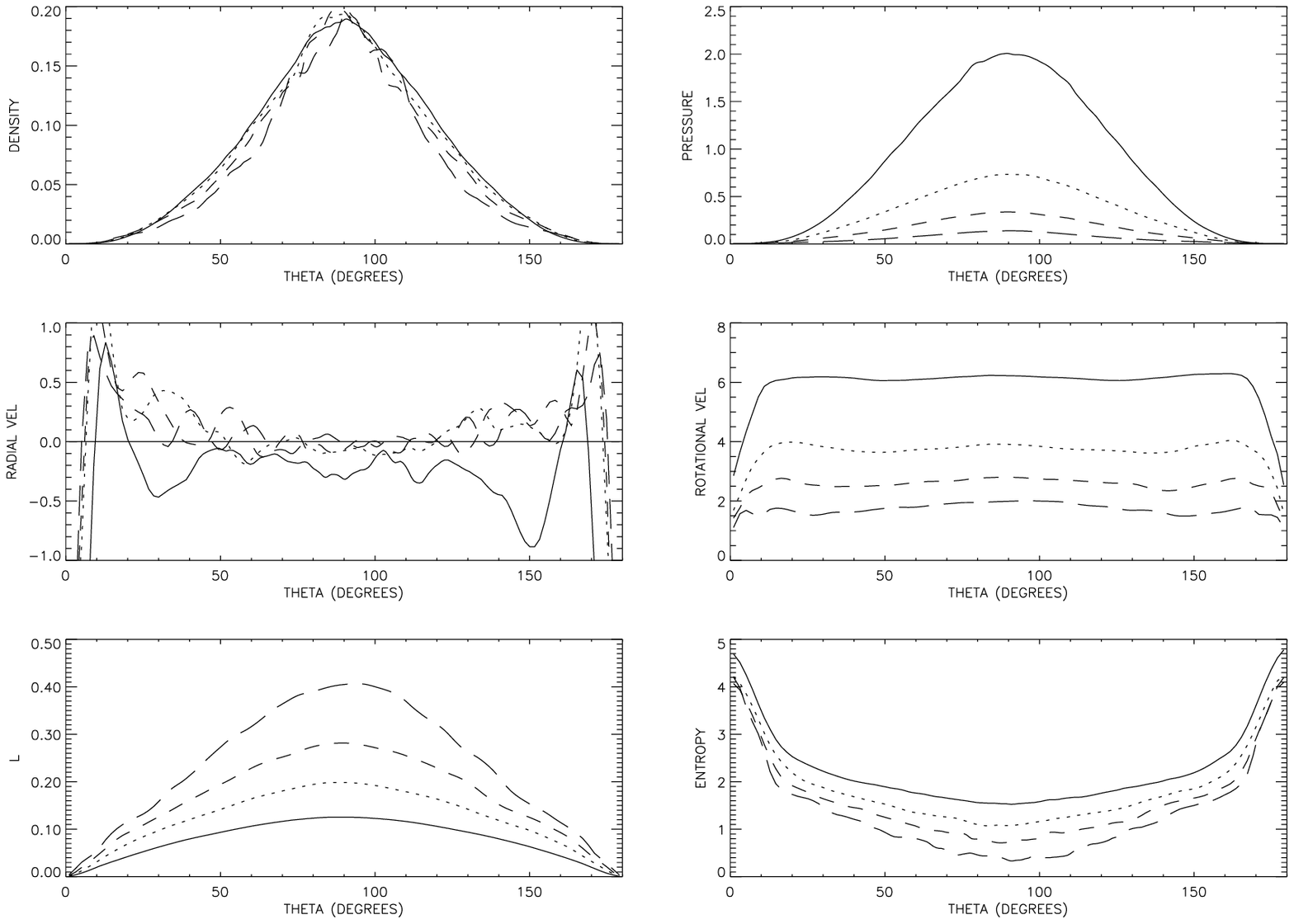}}
\end{picture}
\caption{Angular profiles of a variety of time-averaged
variables from the fiducial model (Run B) at radial positions of
$r=2R_{B}$ (solid line),
$r=5R_{B}$ (dotted line), $r=10R_{B}$ (dashed line), and $r=20R_{B}$
(long-dashed line).}
\end{figure*}

A plot of the time-averaged angular velocity reveals
$v_{\theta}$ is negative (positive) within 50 degrees of $\theta=0$
($\theta=\pi$), and nearly zero elsewhere.  The angular and radial pattern
of $v_{r}$ and
$v_{\theta}$ indicate that for Run B, there is a weak global
circulation pattern consisting of slow radial infall near the equator,
and faster polar outflow above $\theta \approx 45$ degrees.
The average velocity of the circulation is much smaller than the
instantaneous fluctuations, especially near the equator.

The increase of  $\dot{M}_{\rm in}$ and $\dot{M}_{\rm out}$ with radius
indicates most of the mass which flows inward near the equatorial plane
eventually flows outward towards the poles.  Fundamentally, this result
seems to be a consequence of strong convection in the inner regions
which, when combined with diffusion on the smallest scales, serves to
mix the specific entropy of the fluid.  Thus, outflowing material has
only a slightly larger specific entropy than the rest of the material,
so that it is very inefficient at carrying away energy liberated by the
infalling gas (BB).  The result is that a large amount of outflowing
mass is required to remove the energy liberated by the small accretion
rate.

To confirm that our solution is independent of the location of the
inner boundary, we have calculated the evolution of two additional
models, one with $R_{B}=0.005$ (Run C), and the other with $R_{B}=0.02$
(Run D), but otherwise identical to Run A.  The final, time-averaged
$\dot{M}_{\rm acc}$ through the inner boundary measured from these runs is
listed in Table~1.  The results are consistent with $\dot{M}_{\rm acc}
\propto r$.  Since for the particular accretion flow established in our
simulations the viscous accretion rate is proportional to $r$, this
result again indicates that the net mass accretion rate is determined
locally at the surface of the central object.  The details of the flow
in Runs C and D are similar to Run A, we find the same radial and angular
structure of the time-averaged flow as is shown in Figs. 4-7.
In particular, the outflowing
material is still bound, and has the same total energy as in Run A, and
therefore stagnates at roughly the same radial position as the flow
shown in the last panel of Fig.~1.  Making the inner boundary smaller
does not lead to the production of a more energetic outflow.  Of
course, varying the location of the inner boundary is equivalent to
moving the initial location of the torus either closer to or farther
from the central point mass.  Since most of the gravitational energy is
liberated near the central object, we should not expect substantial
changes in the flow if the torus starts at $200R_{B}$ as opposed to
$100R_{B}$, as is indeed observed.

We have studied the direction of the time-averaged advective flux of
mass, energy, and angular momentum, and the flux of $L$ transported by
shear stresses in Run B.  We find the first three are nearly identical,
thus in Figure 8 we show a vector plot of only $\dot{M}=2\pi r^{2} \rho
v_r \sin \theta$
for the same 139 data files as used to construct Figs.  4-7.  To
further reduce the noise in the data, we have averaged together the fluxes on
either side of the equator.  Even so, the direction of the inflow
does not form smooth streamlines.  Overall, there is a quasi-radial
inflow within $\theta \sim 45$ degrees, and a more nearly radial outflow above
that (although, consistent with the profile of the time-averaged angular
velocities mentioned above, the
mass flux vectors are more polar than equatorial).
Note from Fig. 8 that one could also use the change in sign of the radial
component of the average mass flux vectors to crudely define a surface
to the disc at $\theta \sim 45$ degrees.
Although not plotted, the viscous flux of angular momentum is radial within the
inflowing regions, and away from the poles in the outflow.

\begin{figure*}
\begin{picture}(252,180)
\put(0,0){\includegraphics{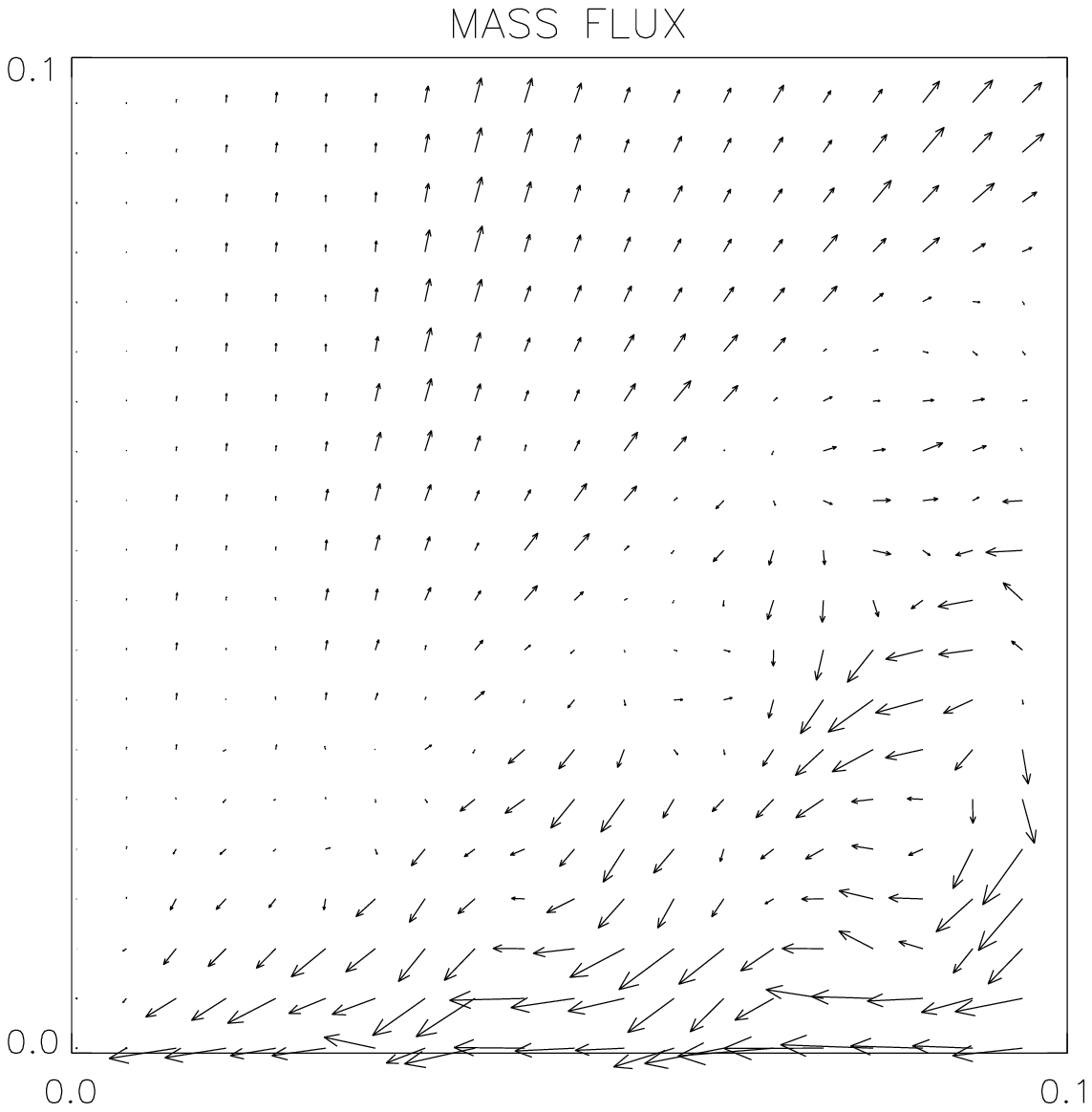}}
\end{picture}
\caption{Vectors of the time-averaged mass flux in the fiducial model (Run B).
The data has also been averaged across the equator and the vectors have been
scaled by $r$.  The maximum vector
length corresponds to a mass flux of magnitude 0.26 (in units where
$G=M=R_0=1$) at $r=0.01$.}
\end{figure*}

The accretion flow which emerges from these calculations must
be driven by angular momentum transport associated with the
anomalous shear stress.  This is clearly demonstrated by setting
the shear stress to zero once the mass accretion rate has reached
a steady-state.  When Run A is continued for another orbit with $\nu=0$,
the mass accretion rate drops by an order of magnitude within 0.1 orbits,
the convective motions die away, and the torus settles into a new
two-dimensional hydrostatic equilibrium with a nearly Keplerian rotation
profile near the equator.

It can also be demonstrated that the thermal heating due to the 
anomalous shear stress plays a fundamental role in producing
convection and mass outflow.  If we repeat the fiducial model with
the heating term (last term in equation 3) removed, we find
a high-velocity, ordered infall is produced with no small-scale
convective motions.

We have studied the effect of varying the adiabatic index $\gamma$ on
the properties of the flow.  Run E listed in Table~1 has $\gamma=4/3$,
but otherwise is identical to the fiducial model Run A (a different
value of $A$ must be chosen in order to keep $M_{t}$ fixed -- this
increases the maximum initial density in the torus).  We find that
apart from being slightly more compact in the equatorial plane (the
opening angle of the accretion being closer to 30 degrees) the detailed
properties of the solution (for instance the radial scaling of
time-averaged variables) do not change significantly compared to Run
A.  The rotational velocity is still very nearly Keplerian, and
$\rho \propto r^{0}$ and $v_{r} \propto r^{-1}$.  Most
significantly, we still find
$\dot{M}_{\rm in} \propto r$ in this simulation.
The mass accretion rate through the inner
boundary listed in Table~1 is increased in Run E compared to A;
it is a small fraction of $\dot{M}_{\rm in}(R_0)$ and is consistent
with the local, viscous accretion rate at the inner boundary.

Finally, to study the effect of the shape of the initial torus, we have
computed two models with $d=1.042$, one with $\gamma=5/3$ (Run F) and
one with $\gamma=4/3$ (Run G).  Once again, $M_{t}$ is kept fixed
between these models by choosing an appropriate value for $A$.  The
smaller value of $d$ used in these models gives the torus
in the initial state a cross-sectional diameter about one half the
value in Run A.
We find no qualitative change in the solution compared to Run A.  In
particular, the outflow is driven to the same radial position (but no
farther), indicating that the initial thermal energy is not important
in the outflow.  In both cases, the flow is dominated by convection in the
inner regions, with time-averaged radial profiles for each variable
that obey the same radial scalings as found for Run A.
The $\gamma=5/3$ (Run F) model has a steady-state mass
accretion rate identical to Run A, while in the $\gamma=4/3$ (Run G)
model the accretion rate is increased by the same amount as Run E.

\subsection{Effect of varying the amplitude of the Shear Stress}

We have calculated the evolution of two models in which $\nu =
10^{-3}\rho$ and $\nu = 10^{-1}\rho$ (i.e., the shear stress is either
ten times smaller or larger than in the fiducial model respectively).
These models are labeled as Runs H and I in Table 1.  Each model is
computed at the standard resolution of 64 grid points per decade in
radius.  Together with Run A, they form a set of models which span two
orders of magnitude in amplitude of the anomalous shear stress.

At low values of the shear stress (Run H), the evolution to a steady
accretion flow proceeds much more slowly than in the fiducial model, as
one would expect.  Thus, it takes roughly 20 orbits of the torus for
$\dot{M}_{\rm acc}$ to converge to a steady value.  This value
(final column in Table 1) is roughly fifteen times
smaller than Run A, although for a viscous disc one expects the mass
accretion rate to be directly proportional to $\nu$.  The discrepancy
can be attributed to the fact that the density at the equator is about 2/3 the
value in Run A.  The two-dimensional structure of the time-averaged
flow in Run H is very similar to that shown in Figure 4 for Run A.  The radial
scaling of all variables near the equator also follow the results of
Run A closely, i.e. $\rho\propto r^{0}$, $P \propto r^{-1}$,
$v_{\phi} \propto r^{-1/2}$, and $v_{r} \propto r^{-1}$.  The ratio of
$v_{r}/v_{\phi} \sim 10^{-3}$ at $R_{0}$, slightly smaller than Run A.
Once again, the inner regions of the flow are dominated by convective
motions which drive a significant outward mass flux that carries away most of
the energy liberated by accretion.  The mass inflow rate
once again scales as $\dot{M}_{\rm in} \propto r$.

At high values of the shear stress (Run I), the evolution of the torus
becomes very dynamic.  In fact, the initial infall of material from
the torus becomes nearly supersonic, reaching the rotation axis after
only 0.15 orbits.  This rapid infall creates a high pressure as material
reflects off the rotation axis, resulting in the ejection of some matter
along the poles.  Convective motions begin within $r<0.1$ after 0.2 orbits,
at which point the flow attempts to settle into the same equatorial-inflow
polar-outflow solution observed in Run A.  However, the mass accretion
rate is marked by large amplitude fluctuations throughout the evolution
of Run I, leading to a time-averaged value only 3 times larger than Run A.
The time-averaged radial scaling of all variables shows the same
dependences on $r$ for all variables as observed in Run A for $r>0.1$,
however for $r<0.1$ some variables diverge from simple power-law
behavior.  For example, both the rotational and radial
velocity becomes flat for $r<0.04$.  At the inner edge,
the ratio of $v_{r}/v_{\phi} \approx 0.5$, and $v_{\phi}/C_{s} \approx 0.5$
(note for Run A this latter ratio $\sim 1.2$).  These ratios
indicate the flow has deviated from strictly Keplerian rotation, and
that radial pressure gradients are playing an important role in the
radial structure of the flow.  Despite the fact that
the flow at such large values of the shear stress
(the ``effective-$\alpha$" of Run I at the inner boundary is $\alpha_{eff} =
\nu \Omega/C_{s}^{2} \approx 0.1$) is qualitatively different
in the inner regions compared to the fiducial model,
we still observe very low mass accretion rates
through the inner boundary compared to the mass inflow rate at
large radii $\dot{M}_{\rm in}(R_0)$.

Because the violent initial collapse of the torus may be interfering
with the establishment of a steady accretion flow, we have also
calculated a model which begins from the final state of Run A, but with
a shear stress which is slowly increased until $\nu=10^{-1}\rho$.  That
is, starting from time 4.325 in Run A, the model is evolved for roughly
one orbit with $\nu=2\times10^{-2}\rho$, one orbit with
$\nu=5\times10^{-2}\rho$, and finally one orbit with
$\nu=10^{-1}\rho$.  The final time-averaged mass accretion rate in the
model (Run I$^\prime$) given in Table 1 is only a factor of 2.4 larger
than the initial accretion rate measured in Run A.  In fact, almost no
change in the accretion rate is observed after the final increase in
$\nu$, indicating that the mass accretion rate is limited at large
values of the stress.  The radial profiles of each variable in Run
I$^\prime$ follow power laws over a wider range in radii, with $v_{r}$
and $v_{\phi}$ becoming flat only over $r<0.02$.  The ratio of
$v_{r}/v_{\phi} \approx 0.3$ at the inner edge, while $v_{\phi}/C_{s}
\approx 0.7$.  The flow in Run I$^\prime$ is again marked by large
amplitude fluctuations in the mass accretion rate.  Thus, we conclude
that at high values of the shear stress, the flow deviates from a
rotationally supported disc due to radial pressure support.
However, we emphasize that strong mass outflow and
convection are still important.

Finally, we note that IA99 have also studied the effect of varying the
amplitude of the shear stress on non-radiative accretion flows.  At low
values of $\nu$, they find strong, small-scale convective eddies
dominate the flow, in accordance with the results presented here.  At
large $\nu$ (much larger than the largest value we have studied here),
they find the flow becomes more laminar, and produces a bipolar outflow
structure.  Qualitatively, our results are in agreement, although there
is an indication of stronger convection in our solutions at large
$\nu$.  It is possible that the poloidal shear stress used by IA99 may
partially suppress convection in this case.

\subsection{Effect of Varying the Form of the Shear Stress}

The last two models listed in Table 1 have been computed with
entirely different forms for the anomalous shear stress than Runs
A through I discussed above.  Run J uses a spatially constant
stress, whereas Run K uses a stress proportional to $r^{1/2}$.
In both models, the amplitude is normalized to give roughly
the same stress at $R_{0}$ as in Run A (about $10^{-3}$ since
$\rho \sim 0.2$ at late times).  We use Runs J and K to investigate the effect
of changing the form (rather than amplitude) of the stress on the
resulting flow.

Since, once a steady-state is achieved in the inner regions, the
accretion flow in which $\nu \propto \rho$ has constant density along
the equatorial plane (leading to a constant shear stress with radius),
we expect the flow in Run J to be similar to Run A.  Indeed, we find
that in Run J the mass accretion rate eventually saturates at
approximately the same value as Run A.  The flow is once again
dominated by large amplitude convection motions in the inner regions.
Most importantly, the time-averaged two dimensional and radial
structure of the flow is similar to that found in Run A.  In
particular, we again find $\rho \propto r^{0}$, $v_{r} \propto
r^{-1}$, and $\dot{M}_{\rm in} \propto r$, while the ratio $v_{r}/v_{\phi}
\sim 10^{-2.5}$ at $R=R_{0}$.  Perhaps the only significance of this
result is that although the shear stress in the outflowing material is
greatly increased in Run J compared to Run A (because of the variation
in density with $\theta$) the details of the inflow-outflow solution
have not been affected.  Both the same amplitude for the mass accretion
rate and the same radial scaling of the time-averaged solution are
obtained.

In Run K the shear stress is given a fixed radial
scaling which differs from the profile of the stress eventually
established in Runs A and J; thus we might expect a quite different
flow pattern to result.
The mass accretion rate eventually saturates in Run K after 10 orbits
at a value consistent with Run A.
Figure 9 plots two-dimensional contours of time-averaged variables in
Run K constructed from 33 equally spaced data files between orbits
12.0 and 13.6.  The general shape of the contours (from concave to
convex with respect to an outward norm at the equator) follows
the same pattern as observed in Run B (Fig. 4).  In particular,
the systematic trend in the shapes indicates that once again,
strong convection drives the time-averaged flow into a marginally
stable state.  However, there are important differences in the solution;
in particular the density at the midplane is now clearly decreasing
with radius, while the Bernoulli function is negative over nearly
the entire domain.
The instantaneous structure of each variable at any time after
the flow has reached a steady accretion rate is close to that
shown in Figure 3 for Run B -- namely large amplitude fluctuations
around the mean values shown in Figure 9 associated with strong convection.

\begin{figure*}
\begin{picture}(504,360)
\put(0,0){\includegraphics{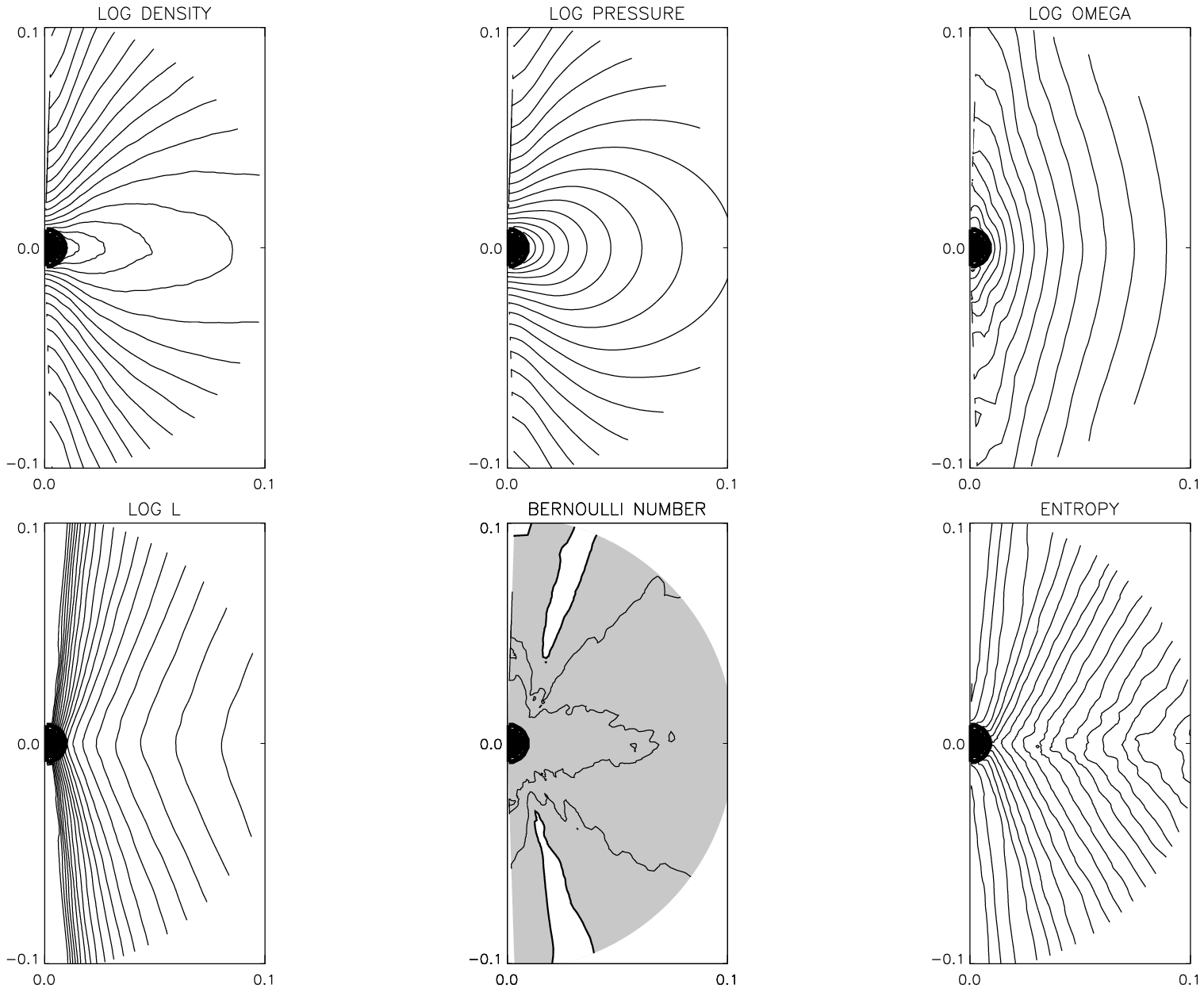}}
\end{picture}
\caption{Two-dimensional structure of a variety of time-averaged
quantities from the $\nu=10^{-3}r^{1/2}$ model (Run K).  The minimum and maximum
of $\log \rho$ are -2.1 and 0.23, of $\log P$ are -1.5 and 1.6,
of $\log \Omega$ are 1.4 and 3.0, of $\log L$ are -2 and -0.59
and of $S$ are -0.25 and 0.29.  For these variables, twenty equally spaced
contour levels are used.  Only three contours between -2 and 0 are
used for the plot of $B$ (although the minimum and maximum are -16 and 0.51
respectively), with shaded regions denoting $B<0$.}
\end{figure*}

Figure 10 plots the radial structure of the flow in Run K averaged
between angles of 85 and 95 degrees, using the same time averaged data
as used to construct Figure 9.  Once again, each variable is seen to be
best fit by a power-law scaling with radius, with $\rho \propto
r^{-1/2}$, $P \propto r^{-3/2}$, $v_{\phi} \propto r^{-1/2}$,
and $v_{r} \propto r^{-1/2}$.  Note the scaling of $\rho$, $P$, and
$v_{r}$ differ from that found for Run B; in this case the solution is
indeed self-similar.  The radial scaling of $C_{s}^{2}$ follows
equation 8 reasonably well.

\begin{figure*}
\begin{picture}(504,360)
\put(0,0){\includegraphics{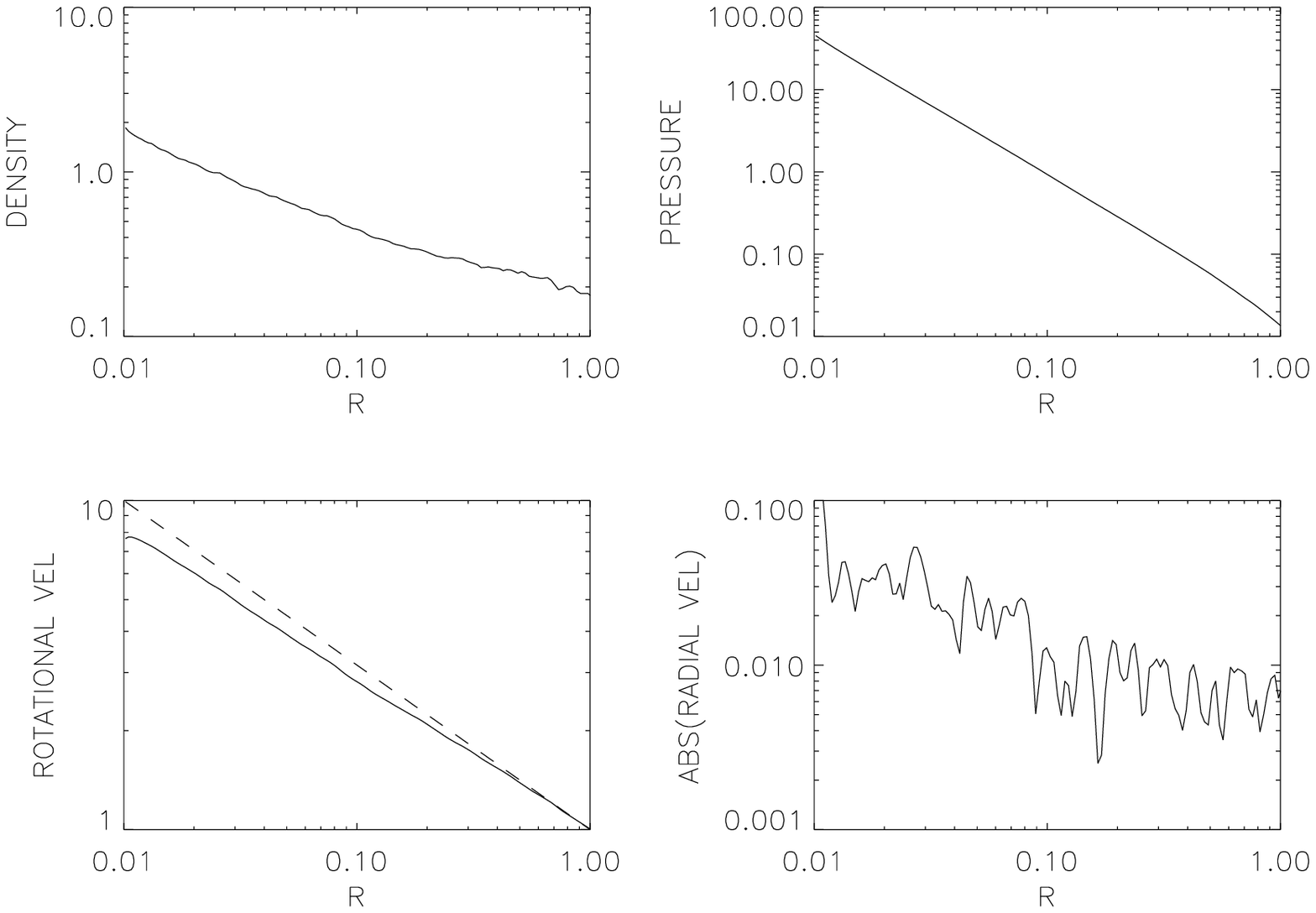}}
\end{picture}
\caption{Radial scaling of some time-averaged
quantities from Run K.  The solution is averaged over
angle between $\theta=85$ and 95 degrees to construct each plot.
The dashed line in the plot of $v_{\phi}$ denotes Keplerian rotation
at the equator.}
\end{figure*}

\begin{figure*}
\begin{picture}(504,360)
\put(0,0){\includegraphics{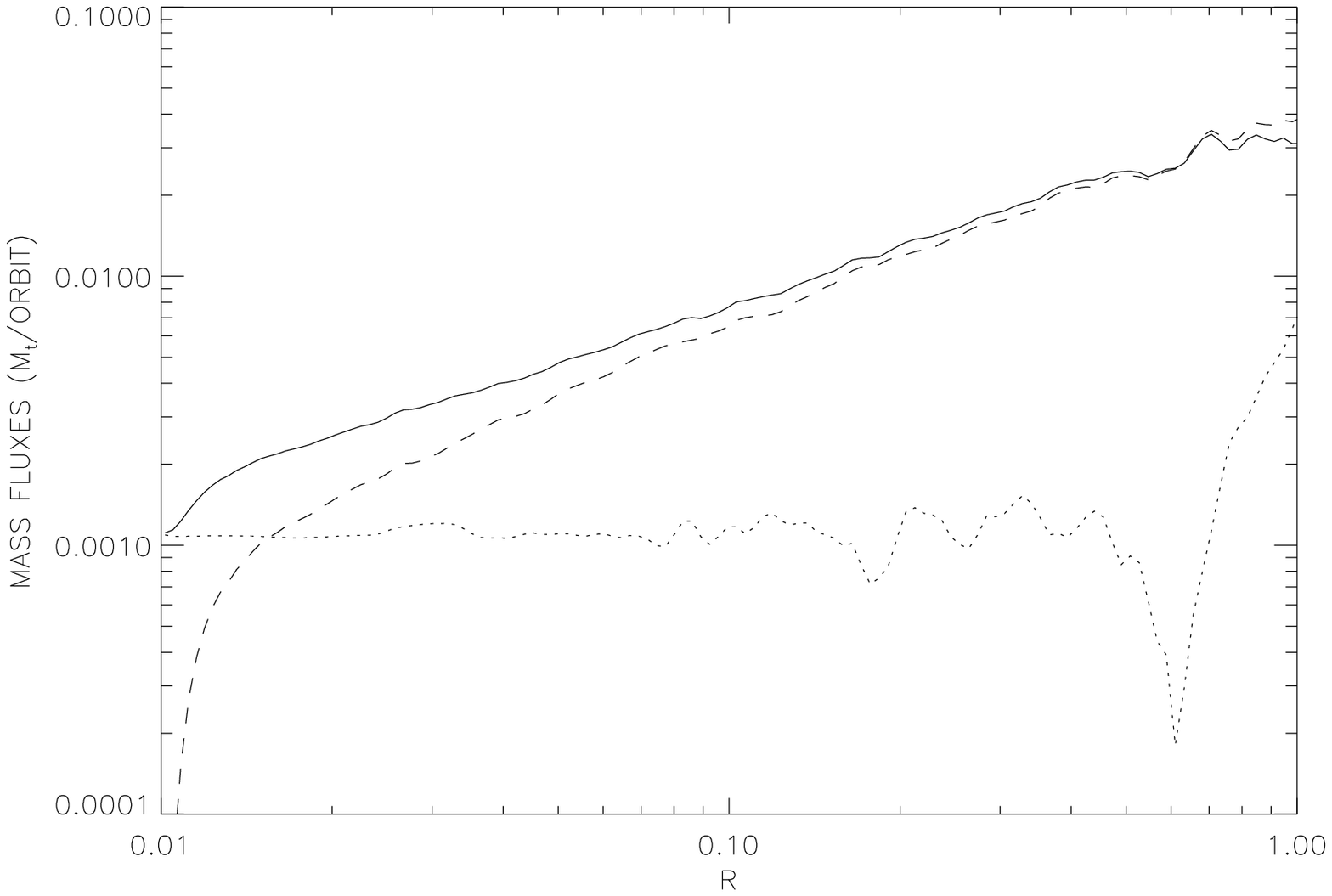}}
\end{picture}
\caption{Radial scaling of the total mass inflow rate $\dot{M}_{\rm in}$
(solid line), total mass outflow rate $\dot{M}_{\rm out}$ (dashed line), and
the net mass accretion rate $\dot{M}_{\rm acc}$ (dotted line)
in Run K.  These fluxes are defined by equations 8, 10, and 11
in the text.}
\end{figure*}

Figure 11 plots $\dot{M}_{\rm in}$, $\dot{M}_{\rm out}$, and $\dot{M}_{\rm acc}$ for
Run K using the same time-averaging
as in Figure 9.  While both the mass inflow and outflow rates increase
rapidly with $r$, they do so less steeply than in the fiducial model
(c.f. Fig. 6): here $\dot{M}_{\rm in} \propto r^{3/4}$ roughly.  The net mass
accretion rate  $\dot{M}_{\rm acc}$ is remarkably constant throughout
the flow, and has a value consistent with the viscous accretion rate
for the properties of the flow near the inner boundary.

Figure~12 plots the angular structure of the time-averaged flow in Run
K at radial positions of $r=0.02$, 0.05, 0.1, and 0.2, again using the
same time-averaging as in Fig. 9.  The angular structure of both the
density and pressure shows smooth variation from the poles to equator,
with the amplitude of both dropping with radius.  Unlike the profiles
of the radial velocity for the fiducial model Fig. 7, $v_{r}$ averages
to almost zero over most angles in Run K.  Like the radial velocity, the
time-averaged angular velocity is almost zero over most angles.  As in
Fig. 7, the rotational velocity $v_{\phi}$ is nearly constant with
$\theta$ at every radial position, so that $L$ varies as $\sin
\theta$.  Note that the profile at $r=2R_{B}$ (solid line each panel)
may be affected by numerical resolution near the poles:  Run K is
computed at only half the resolution of Run B.
Comparison of the structure of the solution in Runs A and B (the
same model computed at different resolutions) indicate that within
$r=2R_{B}$ material can fill in the gap near the poles in the
lower resolution simulations.  The radial and angular profiles
of $v_{r}$ and $v_{\theta}$ indicate there is no global circulation
pattern in Run K, instead inflow and outflow are evenly distributed throughout
the disc.  In this respect, the flow is closer in analogy to a rapidly
rotating star than an accretion disc, in that thermal energy is produced
deep in the interior and is convected outwards at all angles.

\begin{figure*}
\begin{picture}(504,360)
\put(0,0){\includegraphics{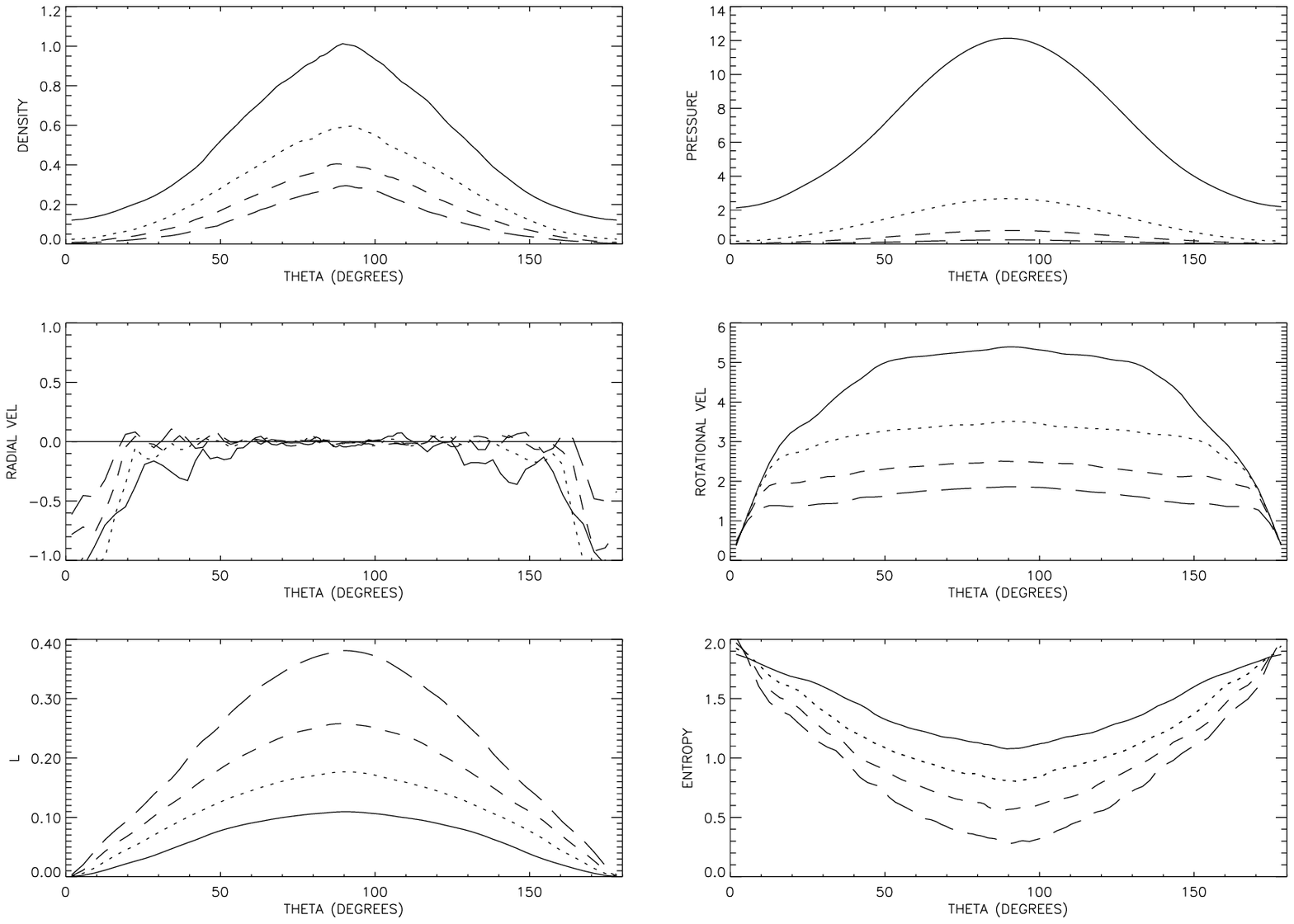}}
\end{picture}
\caption{Angular profiles of a variety of time-averaged
variables from Run K at radial positions of $r=2R_{B}$ (solid line),
$r=5R_{B}$ (dotted line), $r=10R_{B}$ (dashed line), and $r=20R_{B}$
(long-dashed line).}
\end{figure*}

\section{Discussion}

Our major finding is that, independent of the equation of state of the
gas, and independent of the magnitude and functional form of the shear
stress, the angle-integrated mass inflow rate strongly increases with radius,
$\dot{M}_{\rm in} \propto r$ for Runs A-J, and $\dot{M}_{\rm in} \propto r^{3/4}$
for Run K.  In every case, the mass inflow
is balanced by a nearly equal mass outflow, so that
the actual accretion rate onto the central object $\dot{M}_{\rm acc}(R_{B})$ is
only a fraction of the canonical accretion rate $\dot{M}_{\rm in}(R_{0})$ which
should be generated by the torus at radius $R_0$ for the given shear stress,
\begin{equation}
\dot{M}_{\rm acc}(R_{B})/\dot{M}_{\rm in}(R_{0}) \sim (R_{B}/R_{0})^{q}
\end{equation}
where $q=1$ (Runs A-J) or $q=3/4$ (Run K).
For most of the models presented here, $R_{B}/R_{0} = 0.01$, although we
find this scaling also applies over a range of this ratio.
The net mass accretion rate through the flow
is constant with radius (as it must be for a steady state) and is 
consistent with the local viscous accretion
rate given by the properties of the flow near the inner boundary.

Although this result is independent of the nature of the shear stress,
the detailed properties of the solution do in fact change according to the
functional form of the stress.  For example, models in which the shear
stress per unit volume $\nu$ is constant give $\rho \propto r^{0}$ and
$v_{r} \propto r^{-1}$, whereas models in which $\nu \propto r^{1/2}$
give $\rho \propto r^{-1/2}$ and $v_{r} \propto r^{-1/2}$.  In fact,
these scalings can be understood directly from the equations of motion,
using the known properties of the solutions.  For example, using the
fact that the time-averaged rotational velocity is nearly Keplerian
($v_{\phi} = r^{-1/2}$) and symmetric across the equatorial
plane, the $\phi$-component of the momentum equation in spherical polar
coordinates can be written (assuming steady state) as
\begin{equation}
\left( \frac{3}{2} \right)^{2} \frac{\nu}{r^{2}} +
\left( \frac{3}{2} \right)\frac{\nu}{r}\frac{1}{\rho}\frac{\partial\rho}{\partial r}
+ \left( \frac{3}{2} \right) \frac{1}{r} \frac{\partial\nu}{\partial r}
+ \frac{v_{r}}{2r} = 0.
\end{equation}
For this equation to hold over many orders of magnitude in radius, all
terms must scale with $r$ in the same way.  Thus, one sees immediately
that for $\nu=$ constant, $v_{r} \propto r^{-1}$, whereas for $\nu
\propto r^{1/2}$, $v_{r} \propto r^{-1/2}$.

As discussed in \S3, the solutions computed here also fit the condition
that the time-averaged Bernoulli function $B$ is approximately zero in
the equatorial plane.  For nearly Keplerian rotation with negligible
polar velocities, this implies $P/\rho = (\gamma-1)/2\gamma r$
(equation 9).  Using this condition in the energy equation, combined
with the continuity equation and the angular momentum equation given
above, it is possible to predict not only the radial scalings, but also
the amplitudes of the flow variables.  We find such predictions are
only within a factor of a few of the measured time-averaged flow
variables plotted in Figs. 5 and 10.  It is possible this discrepancy
arises because of the difference between constructing nonlinear functions of
time-averaged variables versus the time-average of the functions themselves.

Because almost all of the inflowing material at any particular radius
in the flow is balanced by a nearly equal and opposite
outflow which carries away the energy liberated by
accretion, the solutions found here may be considered as the
``maximally inefficient" accretion solution,
one limit of the ADIOS solutions discussed by BB.  We emphasize,
however, that we do not see powerful, unbound winds in our flows,
rather only slow, outflow associated with convective eddies.  Our flows are
inefficient because vigorous convection in the flow quickly
equalizes the specific entropy of all the gas, thus in order for a
small fraction of the gas to lose energy and accrete inwards, a large
fraction must carry off that energy in a slow outflow.  There may be
some physical effects (such as torques from strong magnetic fields)
that can drive much faster outflows that have a much higher specific
energy.  In this case, it is possible that a smaller fraction of the
inflowing material will be lost to the outflow, and therefore $\dot{M}_{\rm in}$
may not be as steep as $r$ (BB).  Equally important, magnetic fields
will regulate angular momentum transport through the magnetorotational
instability, and thus in MHD the gas does not require an {\em ad hoc}
anomalous shear stress to accrete.  Fully MHD calculations that can
capture both of these effects are underway.

Besides the neglect of magnetic fields, our solutions may also be
limited by the assumption of axisymmetry.  In a fully three-dimensional
flow, angular momentum can be carried outwards by global waves (for
example, our initial state is unstable to a non-axisymmetric global
instability, Papaloizou \& Pringle 1984), and perhaps through the action of radial convection.
However, we do not expect these effects to change the qualitative
nature of the axisymmetric solutions found here.  Fully
three-dimensional simulations are also underway to test this
expectation.

Of course, having $\dot{M}_{\rm in} \propto r$ has important implications for
the observations of accreting compact objects (BB; Quataert \& Narayan
1999) -- the apparent lack of
high-energy emission from such objects may be a result of the
vanishingly small accretion rate in the inner regions.  It is equally
interesting to determine the long term evolution of matter accreting
onto compact objects through the non-radiative accretion flow solutions
presented here. 
The ultimate fate of the outflowing material
depends on whether it is able to cool at large radii and thereby rejoin
the accretion flow (except for the small fraction needed to carry away
the angular momentum).  It is feasible that such a recycling process
may lead to an increase in the density in the inflowing material.
Potentially, this can lead to an increase in the radiative efficiency,
and could make the flow transition to a thin disc with a much larger
accretion rate.  Most of the mass accreted by the central object may be
accumulated through such episodic bursts of accretion.
We must express
a strong word of caution at this point, however.  We have presented
solutions over two orders of magnitude in radius for at most twenty orbits of
the outer regions.  Such evolutionary effects will require
understanding the solutions over five or more orders of magnitude in
radius for thousands of orbits of the outer regions.

\section{Conclusions}

We have investigated the properties of non-radiative rotating accretion
flows by carrying out a set of numerical two-dimensional (axially
symmetric) hydrodynamical experiments using a simple starting
configuration and a set of well-defined boundary conditions.  We start
with all the mass rotating in a torus in hydrodynamic equilibrium at
radius $R_0=1$.  We allow small (in the sense that the induced flow
velocities are small compared to the rotational velocities) anomalous
shear stresses to act on the flow.  We set a purely accreting inner
boundary at (for most of our models) a radius of $R_{B}=0.01$.  All
energy deposited in the fluid by shear stresses is retained by the
fluid, and not radiated.  A spherical polar grid with logarithmically
spaced zones is used to resolve the flow over more than two decades in
radius.

We find that in the inner regions, the flow becomes strongly
convective, and in fact convective motions dominate the instantaneous
structure of the flow.  This convection produces outward moving fluid
elements (with high entropy, low density, and an excess of angular
momentum compared to rotation at the local Keplerian value) at all
angles, including near the equator.  
The outflowing material removes
mass, angular momentum, and most importantly energy from the accretion
flow.  However, the strong convection combined with diffusion on small
scales mixes the specific entropy of the
fluid very efficiently, which means that a very large mass loss rate is
required to remove enough energy to allow some fluid to accrete.

Despite the strong time-dependence of the solution, long time-averages
of the dynamical variables show remarkable correspondence to
steady-state solutions.  Two-dimensional contours indicate the
time-averaged flow is marginally stable to the H{\o}iland criterion.
Radial profiles of the variables are all power-laws over nearly the
entire domain.  The rotation profiles are always nearly Keplerian
(except for very large values of the shear stress when, in the inner
regions, pressure gradients become substantial).

Perhaps the most important result of this study is that the mass
accretion rate through the inner boundary is a small fraction of
the angle-integrated mass inflow rate at large radii.  The net
mass accretion rate through the disc is consistent with the viscous
accretion rate appropriate to the conditions in the flow at the surface
of the central object.  Most of the inflowing mass at large radii
eventually becomes part of the outflow driven by convection.
This result is independent of the adiabatic index of the
gas, or the magnitude or form of the anomalous shear stress.

Qualitatively, our results are in agreement with those of Igumenshchev
(1999), who has undertaken a similar study using different numerical
methods, for low values of the shear stress.  If confirmed by more
realistic three-dimensional and MHD calculations, this result has
important implications both for the spectrum and luminosity of
accreting compact objects, and the long term evolution of accretion
flows.

{\bf Acknowledgments:} We thank Roger Blandford for many stimulating
discussions and comments on earlier drafts of this paper, and an
anonymous referee for helpful suggestions.
JS gratefully acknowledges financial support from the Institute of
Astronomy, and from NSF grant AST-9528299 and NASA grant NAG54278.
MB acknowledges support from from NSF grants AST-9529175 and AST-9876887,
and a Guggenheim Fellowship.
This research was supported in part by the National Science
Foundation under Grant No. PHY94-07194.

\end{document}